\documentclass[pre,longbibliography]{revtex4-1}
\usepackage{amsmath,amssymb,latexsym,epsfig,graphics,epsf}
\usepackage{graphics,xcolor}
\usepackage{float}
\newcommand{\br}{\bold r}

\newcommand{\bR}{\bold R}

\newcommand{\be}{\begin{equation}}
\newcommand{\ee}{\end{equation}}
\newcommand{\fig}[1]{Fig.~\ref{#1}}
\newcommand{\Fig}[1]{Figure~\ref{#1}}
\newcommand{\Sec}[1]{Sec.~\ref{#1}}
\newcommand{\eq}[1]{Eq.~(\ref{#1})}

\newcommand{\Sex}{{S}_{\rm ex}}

\begin{document}
\title{Extreme case of density scaling: The Weeks-Chandler-Andersen system at low temperatures}
\author{Eman Attia}\email{attia@ruc.dk}
\author{Jeppe C. Dyre}
\email{dyre@ruc.dk}
\author{Ulf R. Pedersen}
\email{urp@ruc.dk}
\affiliation{``Glass and Time'', IMFUFA, Dept. of Science and Environment, Roskilde University, P. O. Box 260, DK-4000 Roskilde, Denmark}
\date{\today}

\begin{abstract} 
This paper studies numerically the Weeks-Chandler-Andersen (WCA) system, which is shown to obey hidden scale invariance with a density-scaling exponent that varies from below 5 to above 500. This unprecedented variation makes it advantageous to use the fourth-order Runge-Kutta algorithm for tracing out isomorphs. Good isomorph invariance of structure and dynamics is observed over more than three orders of magnitude temperature variation. For all state points studied, the virial potential-energy correlation coefficient and the density-scaling exponent are controlled mainly by the temperature. Based on the assumption of statistically independent pair interactions, a mean-field theory is developed that rationalizes this finding and provides an excellent fit to data at low temperatures and densities. 
\end{abstract}
\maketitle

\section{Introduction}\label{Sec1}

Density scaling is an important experimental discovery of the last 20 years' liquid-state research, which by now has been demonstrated for high-pressure data of hundreds of systems \cite{alb04,rol05,lop12a,adr16}. The crucial insight is that, in order to characterize a thermodynamic state point, the relevant variable supplementing the temperature $T$ is not the pressure $p$, but the number density $\rho\equiv N/V$ (considering $N$ particles in volume $V$) \cite{alb04,rol05,lop12a,adr16,gun11,kiv96}. If $\gamma$ is the so-called density-scaling exponent, plotting data for the dynamics as a function of $\rho^\gamma/T$ results in a collapse \cite{alb04,rol05,lop12a,adr16}. In other words, the dynamics depends on the two variables of the thermodynamic phase diagram only via the single variable $\rho^\gamma/T$. This provides a significant rationalization of data, as well as an important hint for theory development. It should be noted, though, that density scaling does not apply universally; for instance, it usually works better for van der Waals liquids than for hydrogen-bonded liquids \cite{rol05,adr16}.

Some time after these developments were initiated, a framework for density scaling was provided in terms of the isomorph theory \cite{IV,dyr14}, which links density scaling to Rosenfeld's excess-entropy scaling method from 1977 \cite{ros77,dyr18a}. According to isomorph theory, any system with strong correlations between the fixed-volume virial and potential-energy equilibrium fluctuations has curves of invariant structure and dynamics in the thermodynamic phase diagram. These ``isomorphs'' \cite{IV,sch14} are defined as curves of constant excess entropy $\Sex$, which is the entropy minus that of an ideal gas at the same temperature and density ($\Sex<0$ because any system is more ordered than an ideal gas). 

If the potential energy is denoted by $U$ and the virial by $W$, their Pearson correlation coefficient $R$ is defined by

\be\label{R_def}
R
\,=\,\frac{\langle\Delta U\Delta W\rangle}{\sqrt{\langle(\Delta U)^2\rangle\langle(\Delta W)^2\rangle}}\,.
\ee
Here $\Delta$ denotes the deviation from the thermal average and the sharp brackets are canonical ($NVT$) averages. The pragmatic criterion defining ``strong'' correlation is $R>0.9$ \cite{ped08,I}. Systems with strong correlations have good isomorphs, i.e., approximate invariance of structure and dynamics along the configurational adiabats \cite{IV}. Such systems are termed R-simple, signaling the simplification of having an effectively one-dimensional thermodynamic phase diagram in regard to structure and dynamics when these are given in so-called reduced units (see below). Hydrogen-bonded systems usually have $R<0.9$ and are thus not R-simple \cite{ped08}; this explains why density scaling does not apply universally.

Isomorph theory is only rigorously correct in the unrealistic case of an Euler-homogeneous potential-energy function that is realized, for instance, in systems with inverse-power-law (IPL) pair potentials \cite{hey07}. Nevertheless, isomorph-theory predictions apply to a good approximation for many systems, e.g., Lennard-Jones (LJ) type liquids \cite{II,IV,V,yoo19}, the EXP pair-potential system at low temperatures \cite{EXPI,EXPIV}, simple molecular models \cite{ing12b,fra19,kop20}, polydisperse systems \cite{ing15}, crystals \cite{alb14}, nano-confined liquids \cite{ing13a}, polymer-like flexible molecules \cite{vel14}, metals \cite{hum15,fri19}, Yukawa plasmas \cite{vel15,tol19}, etc. 

In some cases, isomorphs are well described by the equation $\rho^\gamma/T=$ Const. with a constant $\gamma$ \cite{sch09}, which as mentioned accounts for density scaling as discussed in most experimental contexts \cite{rol05}. Isomorph theory, however, does not require $\gamma$ to be constant throughout the thermodynamic phase diagram, and $\gamma$ indeed does vary in most simulations \cite{V,ing12a,EXPII,hey19}. The general isomorph-theory definition of the density-scaling exponent $\gamma$ at a given state point \cite{IV,dyr18a} is

\be\label{gamma}
\gamma
\,\equiv\, \left(\frac{\partial\ln T}{\partial\ln \rho}\right)_{\Sex}
\,=\,\frac{\langle\Delta U \Delta W \rangle}{\langle(\Delta U)^2\rangle}\,.
\ee
The second equality gives the statistical-mechanical expression of $\gamma$ in terms of the constant-volume canonical-ensemble fluctuations of potential energy and virial.

The question whether experimental density-scaling exponents are strictly constant throughout the phase diagram has recently come into focus \cite{san19,cas19}. In simulations, isomorphs are in many cases described by the following equation \cite{alb04,boh12,ing12a,dyr14}

\be\label{h}
\frac{h(\rho)}{T}\,=\,\textrm{Const.}
\ee
in which $h(\rho)$ is a function of the density. For the Lennard-Jones (LJ) system, for instance, one has $h(\rho)\propto(\gamma_0/2-1)(\rho/\rho_0)^4-(\gamma_0/2-2)(\rho/\rho_0)^2$ in which $\gamma_0$ is the density-scaling exponent at a reference state point of density $\rho_0$ \cite{boh12,ing12a}. For isomorphs given by \eq{h}, \eq{gamma} implies 

\be\label{hgamma}
\gamma
\,=\,\frac{d\ln h(\rho)}{d\ln\rho}\,.
\ee
We see that unless $h(\rho)$ is a power-law function, the density-scaling exponent depends on the density, though not on the temperature. More generally, $\gamma$ also depends on the temperature \cite{EXPII}. This is the case, for instance, for the LJ system at very high temperatures: for $T\to\infty$ at a fixed density, the LJ system is dominated by the repulsive $r^{-12}$ term of the pair potential, implying that $\gamma$ approaches $12/3=4$ in this limit and that \eq{h} cannot apply.

A likely reason that many experiments are well described by a constant $\gamma$ is the fact that density often does not vary much. As shown by Casalini and coworkers \cite{cas19,ran19}, when extreme pressure is applied, the density-scaling exponent is no longer constant. Although it is now clear that $\gamma$ is not a material constant \cite{san19,cas19}, its variation is as mentioned often insignificant in experiments. This paper gives an example in which $\gamma$ varies dramatically. We present a study of the noted Weeks-Chandler-Andersen (WCA) system \cite{wca,cha83} that 50 years ago introduced the idea of a cutoff at the potential-energy minimum of the LJ system \cite{dek90,bis99,ben04,nas08,ahm09,ben15}. This idea is still very popular and used in various contexts \cite{atr20,daw20,gus20,mir20,nog20,ton20,ued20}.

We show below that the WCA system has strong virial potential-energy correlations and thus is R-simple at typical liquid-state densities. We find that $\gamma$ varies by more than two decades in the investigated part of the phase diagram. In comparison, the LJ system has a density-scaling exponent that varies less than 50\% throughout the phase diagram. To the best of our knowledge, the $\gamma$ variation of the WCA system is much larger than has so far been reported for any system in simulations or experiments. For all state points studied, we find that $\gamma$ depends primarily on the temperature. A mean-field theory is presented that explains this observation and which accounts well for the low-temperature and low-density behavior of the system.

After providing a few technical details in \Sec{Sec2}, the paper starts by presenting the thermodynamic phase diagram with the state points studied numerically (\Sec{Sec3}). The paper's main focus is on three isomorphs, numbered 1-3. Each of these is associated with an isotherm and an isochore, the purpose of which is to put into perspective the isomorph variation of structure and dynamics by comparing to what happens when a similar density/temperature variation is studied, keeping the other variable constant. In \Sec{Sec3} we also give data for the virial potential-energy correlation coefficient $R$ and the density-scaling exponent $\gamma$, demonstrating that all state points studied have strong correlations ($R>0.9$) while $\gamma$ varies from about 5 to above 500. A mean-field theory is developed in \Sec{Sec4a} predicting that $R$ and $\gamma$ both depend primarily on the temperature. Section \ref{Sec4} presents simulations of the structure and dynamics along the isotherms, isochores, and isomorphs. Despite the extreme $\gamma$ variation, which implies that an approximate inverse-power-law description fails entirely, we find good isomorph invariance of the reduced-unit structure and excellent isomorph invariance of the reduced-unit dynamics. Section \Sec{Sec5} gives a brief discussion. Appendix I details the implementation of the fourth-order Runge-Kutta method for tracing out isomorphs and compares its predictions to those of the previously used simple Euler method. Appendix II gives isomorph state-point details.

\section{Model and simulation details}\label{Sec2}

Liquid model systems are often defined in terms of a pair potential $v(r)$. If $r_{ij} =  |\br_i-\br_j|$ is the distance between particles $i$ and $j$, the potential energy $U$ as a function of all particle coordinates $\bR\equiv (\br_1,\br_2,..,\br_N)$ is given by 

\be
U(\bR)\,=\,\sum_{i<j}v(r_{ij})\,.
\ee
We study in this paper the single component Weeks-Chandler-Andersen (WCA) system \cite{wca}, which cuts the standard LJ pair potential at its minimum and subsequently shifts the potential by adding a constant such that the minimum is lifted to zero \cite{wca,han13}. The result is the purely repulsive pair potential given by 

\begin{eqnarray}\label{wca}
v(r)  = 
\begin{cases}
4\varepsilon\,\left[(r/\sigma)^{-12} - (r/\sigma)^{-6} \right]+\varepsilon \,\,&\,\, (r < 2^{1/6}\sigma)\\
0 \,\,&\,\, (r> 2^{1/6}\sigma)
\end{cases}\,.
\end{eqnarray}
Like the LJ pair potential, $v(r)$ involves two parameters: $\sigma$ that reflects the particle radius and $\varepsilon$ that reflects the energy depth of the LJ potential well at its minimum at $r=2^{1/6}\sigma$. 

The WCA system was simulated by Molecular Dynamics (MD) in the canonical ($NVT$) ensemble using the Nos\'e-Hoover thermostat \cite{tildesley}. The simulated system consisted of 4000 particles in a cubic box with periodic boundaries. The simulations were performed using the open-source Roskilde University Molecular Dynamics software (RUMD v3.5) that runs on GPUs (graphics processing units) \cite{RUMD} (http://rumd.org). For updating the system state, the leap-frog algorithm was employed with reduced-unit time step 0.0025. At each state point, a simulation first ran for 25 million time steps for equilibration. This was followed by 50 million time steps for the production run.

The simulations were conducted in the ``reduced'' unit system of isomorph theory in which the energy unit is $e_0\equiv k_BT$, the length unit is $l_0\equiv\rho^{-1/3}$, and the time unit is $t_0\equiv\rho^{-1/3}\sqrt{m/k_BT}$ where $m$ is the particle mass \cite{IV}. A few simulations were also carried out in MD units to check for consistency. Using reduced units in a simulation implies that density and temperature are both equal to unity; thus the state point is changed by varying $\sigma$ and $\varepsilon$, i.e., by changing the pair potential. In contrast, performing simulations in MD units implies putting $\sigma=\varepsilon=1$, i.e., fixing the pair potential and varying $\rho$ and $T$ in order to change the state point. The two methods are mathematically equivalent, of course. Simulating in reduced units is convenient because the time step is then automatically adjusted to take into account the thermal velocity. 

Reduced quantities are generally marked by a tilde, for instance $\tilde{r}\equiv r/l_0=\rho^{1/3}r$. These units are used below for all quantities except for the density and the temperature; thermodynamic state points are reported by giving density and temperature in standard MD units, i.e., $\rho$ is given in units of $\sigma^{-3}$ and $T$ in units of $\varepsilon/k_B$.

\begin{figure}[h]
	\includegraphics[width=6cm]{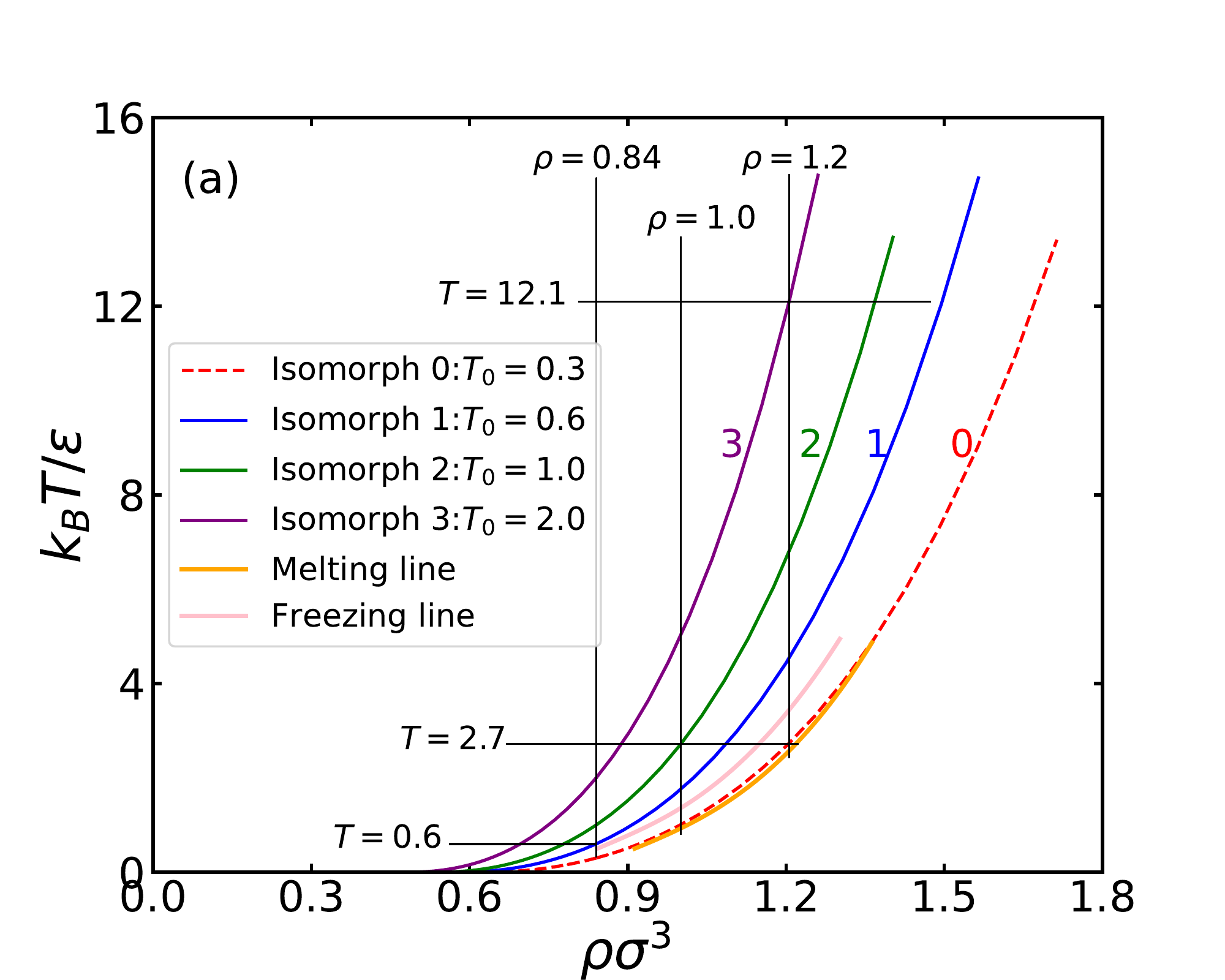}
	\includegraphics[width=6cm]{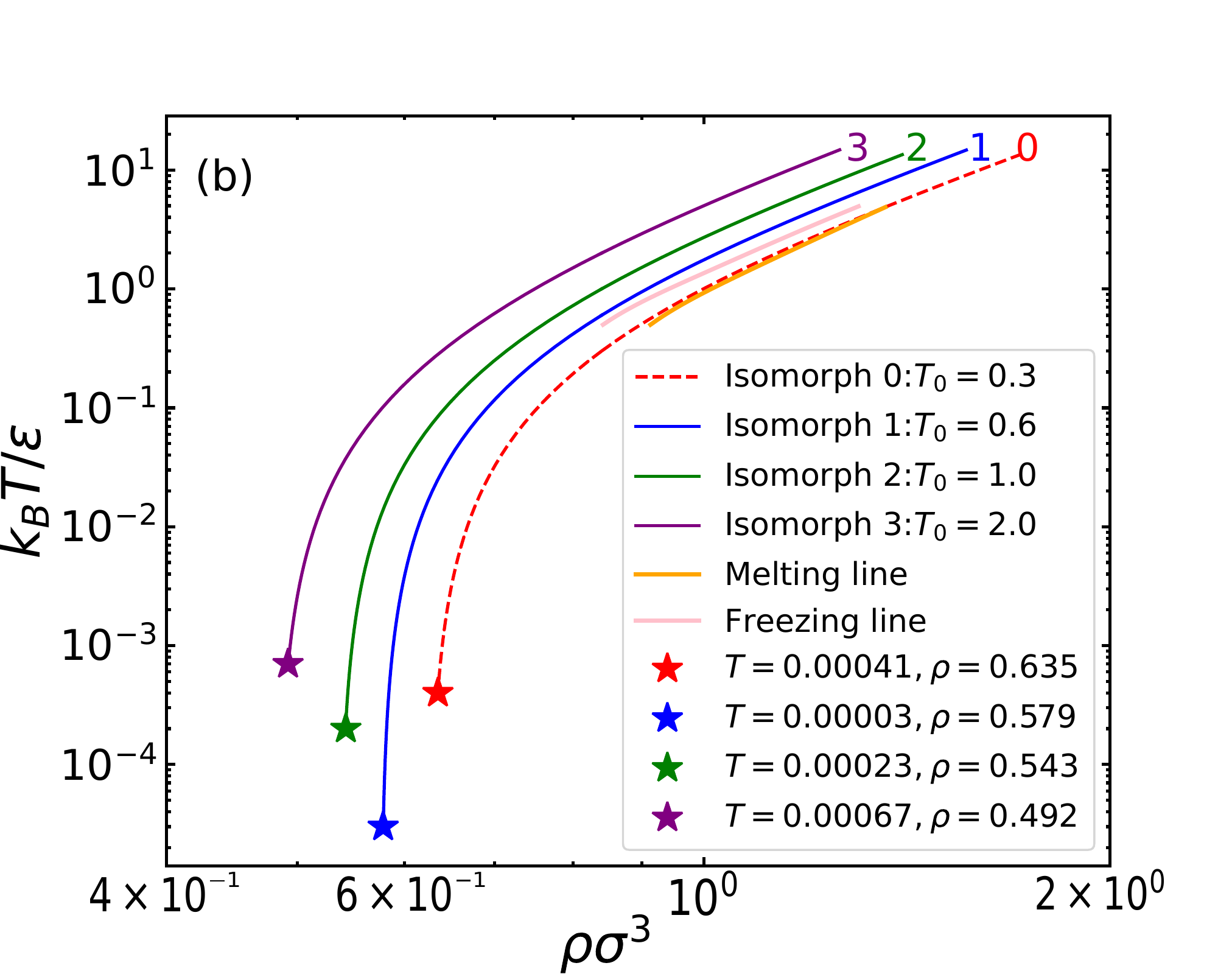}
	\caption{\label{fig1}
		(a) The three isomorphs in focus (denoted 1-3) shown as full curves in the temperature-density thermodynamic phase diagram. Each isomorph was generated as described in the text and in Appendix I, starting from the reference state point ($\rho_0, T_0$) with $\rho_0=0.84$ and $T_0$ equal to 0.6, 1.0, and 2.0, respectively. A fourth isomorph (denoted 0) is marked by the red dashed line and is in the supercooled liquid phase. The horizontal lines are three isotherms and the vertical lines are three isochores, which are studied in order to compare their structure and dynamics variations to those along the isomorphs. The freezing and melting lines are shown as yellow and orange lines, respectively \cite{dek90,ahm09}; note that these are parallel to the isomorphs.
		(b) The four isomorphs shown in a logarithmic temperature-density phase diagram. The slope $\gamma$ (\eq{gamma}) increases significantly as the temperature is lowered along an isomorph. The stars mark the lowest simulated temperature and density on each isomorph; these state points are used in \fig{fig13} below.}
\end{figure}

\section{Simulated state points}\label{Sec3}

\Fig{fig1}(a) shows the thermodynamic phase diagram of the WCA system. The yellow and orange lines to the right are the freezing and melting lines \cite{dek90,ahm09}. The blue, green, and purple lines marked 1, 2, and 3, respectively, are the isomorphs of main focus below, while the red dashed line is a fourth isomorph marked 0, which is in the liquid-solid coexistence region. Note that the freezing and melting lines are both approximate isomorphs \cite{IV,ped16}. 

Each isomorph was traced out starting from a ``reference'' state point of density 0.84. Isomorphs are often identified by integrating \eq{gamma} using the simple first-order Euler integration scheme for density changes of order one percent \cite{IV,V,ing12b}. The extreme variation of $\gamma$ found for the WCA system, however, means that Euler integration can only be used reliably for very small density changes and a more accurate integration scheme is called for. We used the fourth-order Runge-Kutta integration (denoted by RK4) as detailed in Appendix I, where it is demonstrated that RK4 is 10-100 times more computationally efficient that Euler integration for tracing out isomorphs with a given accuracy. Data for selected state points of the four isomorphs are listed in Appendix II.

In order to investigate the degree of isomorph invariance of the reduced-unit structure and dynamics (Sec. \ref{Sec4}), for each isomorph we also performed simulations along an isotherm and an isochore, limiting all simulations to state points in the equilibrium liquid phase, though. \Fig{fig1}(b) shows the isomorphs and the melting and freezing lines in a diagram with logarithmic density and temperature axes. In this diagram the density-scaling exponent $\gamma$ is the isomorph slopes, compare \eq{gamma}, which increases significantly along each isomorph as the density is lowered.

\begin{figure}
	\includegraphics[width=6cm]{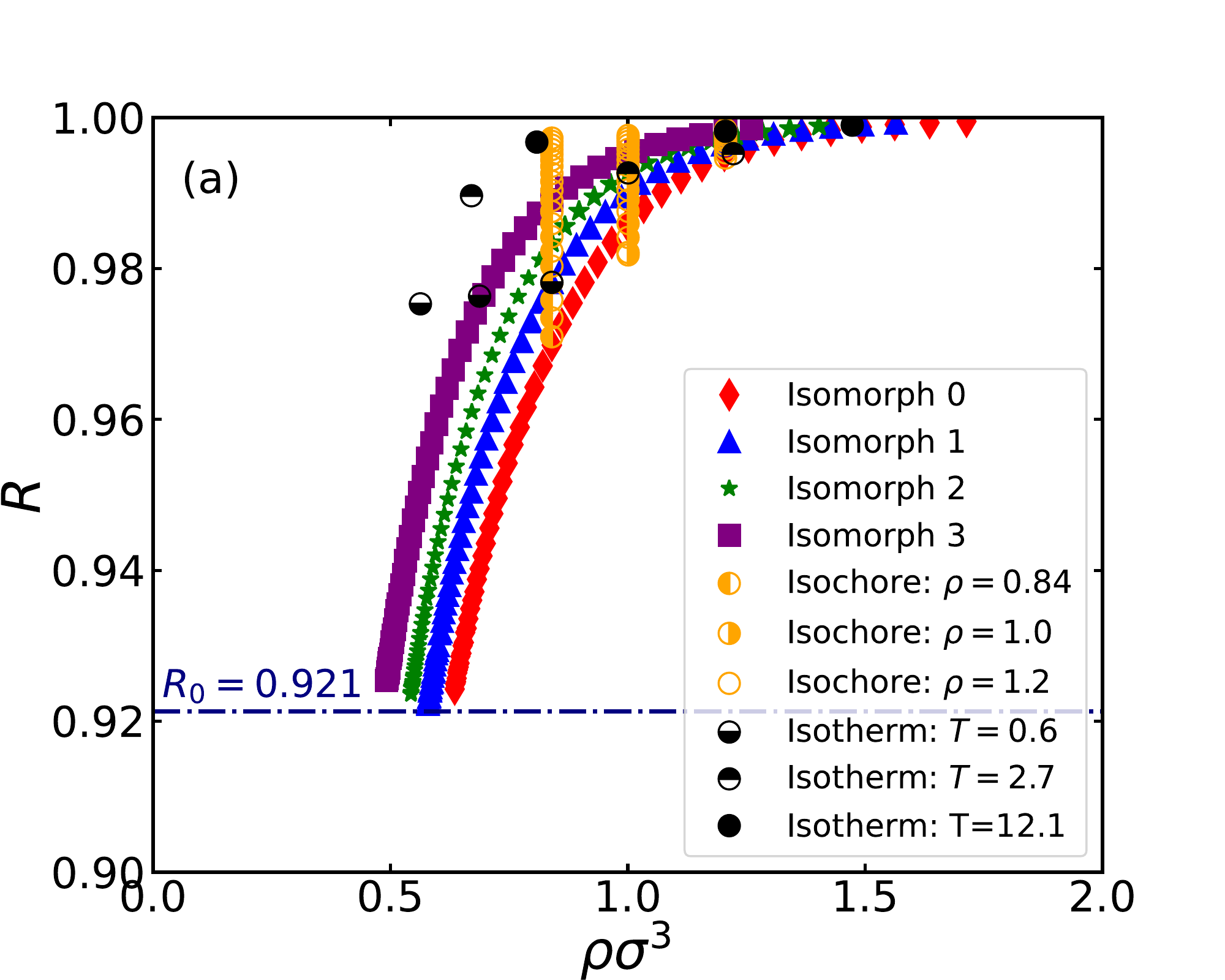}
	\includegraphics[width=6cm]{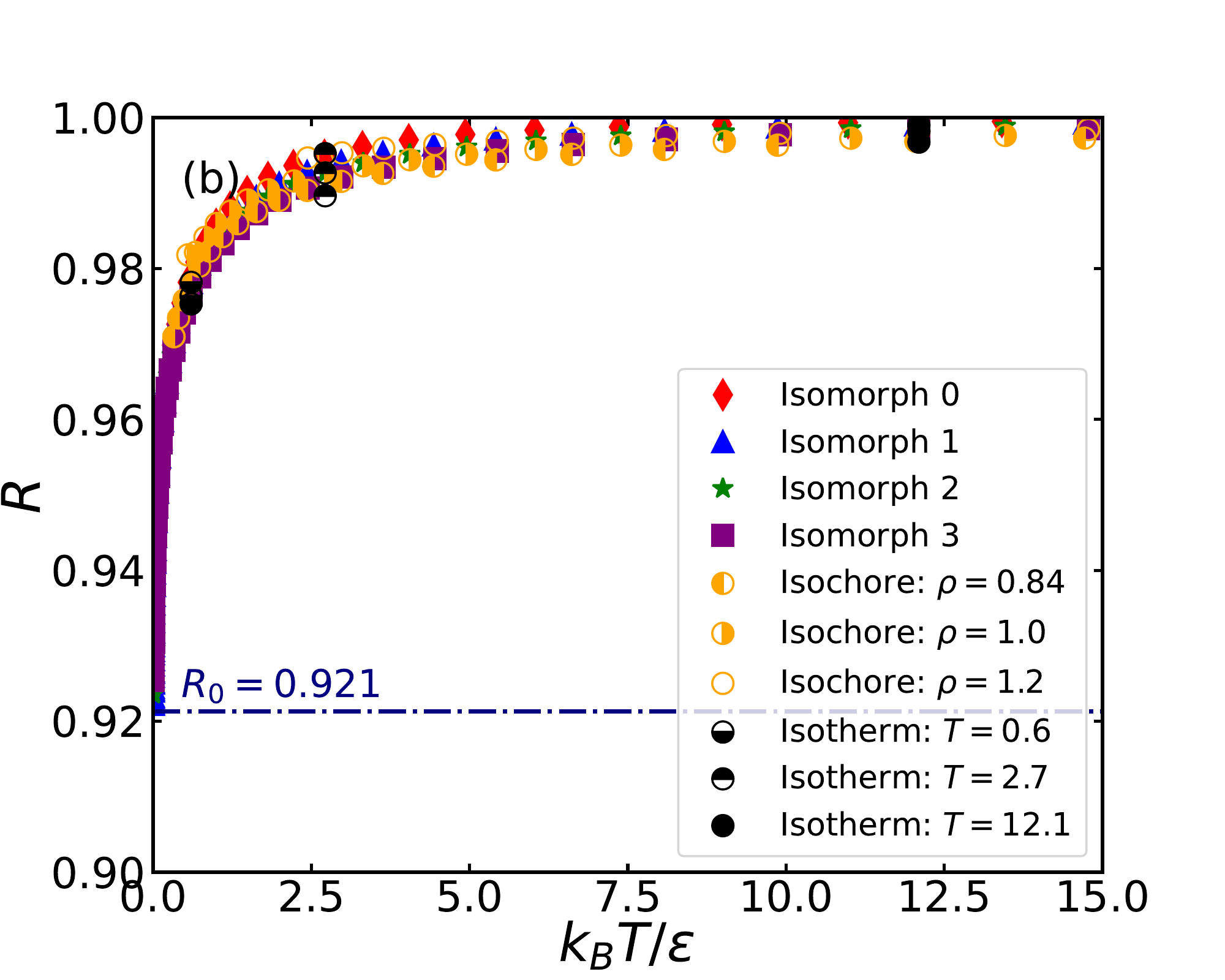}
	\caption{\label{fig2} The virial potential-energy Pearson correlation coefficient $R$ (\eq{R_def}) for all state points studied (\fig{fig1}). There are strong correlations everywhere ($R>0.9$). The horizontal dashed-dotted lines mark the low-temperature, low-density limit of the mean-field-theory prediction, $R=\sqrt{8/3\pi}=0.921$ (\eq{R0}).
		(a) $R$ as a function of the density. 
		(b) $R$ as a function of the temperature. We see that the correlations are mainly controlled by the temperature. }
\end{figure}

A configurational adiabat is an isomorph only for state points with strong virial potential-energy correlations, i.e., when $R\gtrsim 0.9$ at the relevant state points in which $R$ is given by \eq{R_def}. This condition is validated in \fig{fig2}, which shows $R$ for all state points simulated. \Fig{fig2}(a) shows $R$ as a function of the density, while (b) shows $R$ as a function of the temperature. We see that $R$ increases with increasing density and temperature, approaching unity. This reflects the fact that the $(r/\sigma)^{-12}$ term of the pair potential dominates the interactions in these limits and that an inverse-power-law pair potential has $R=1$. An important observation from \fig{fig2} is that strong correlations are maintained even at the lowest densities and temperatures studied. Comparing (a) to (b) reveals that $R$ is primarily controlled by the temperature. This may be understood from a mean-field theory, which assumes that the interactions at low densities are dominated by single-pair interactions (\Sec{Sec4a}).

\begin{figure}[h]
	\includegraphics[width=5.3cm]{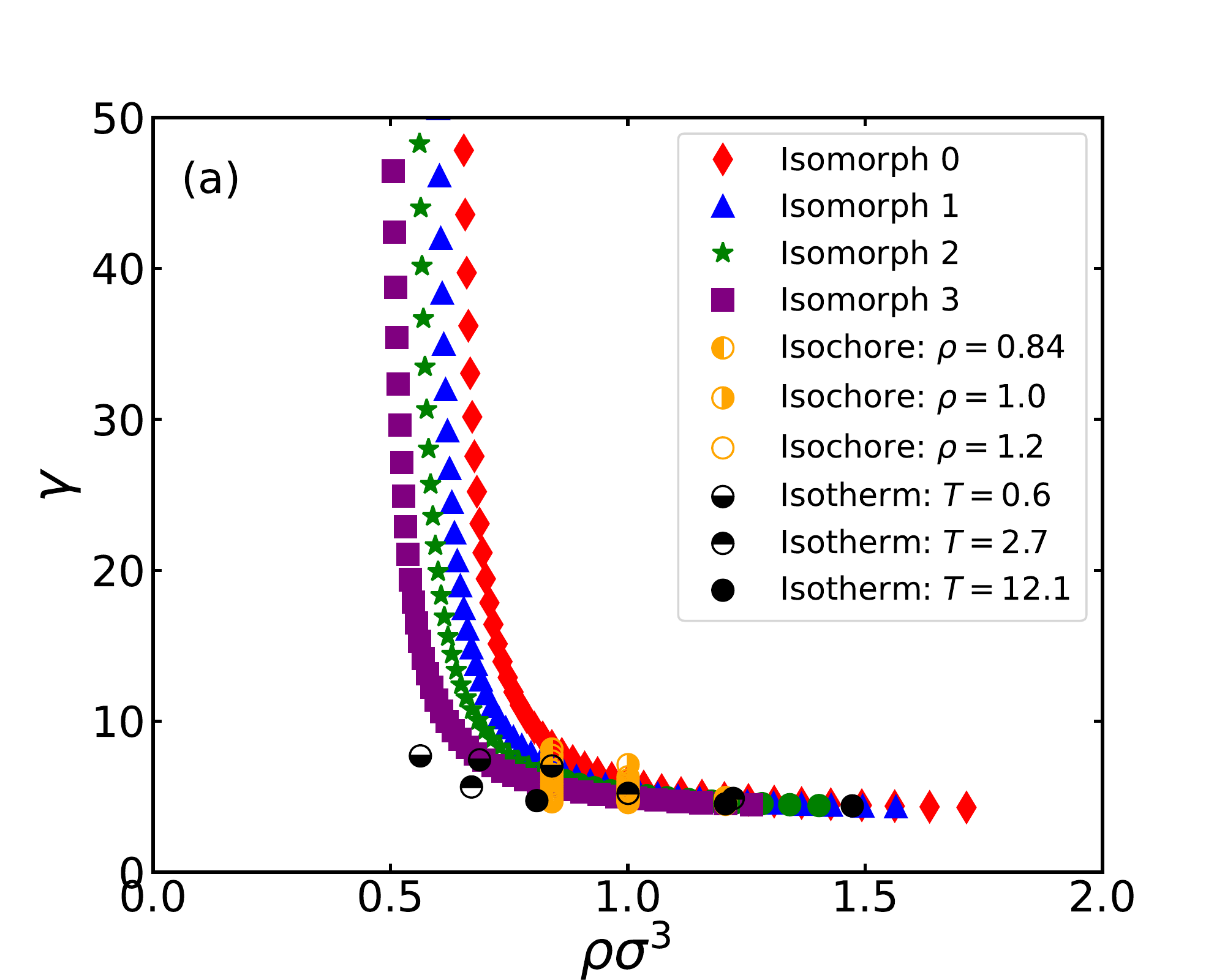}
	\includegraphics[width=5.3cm]{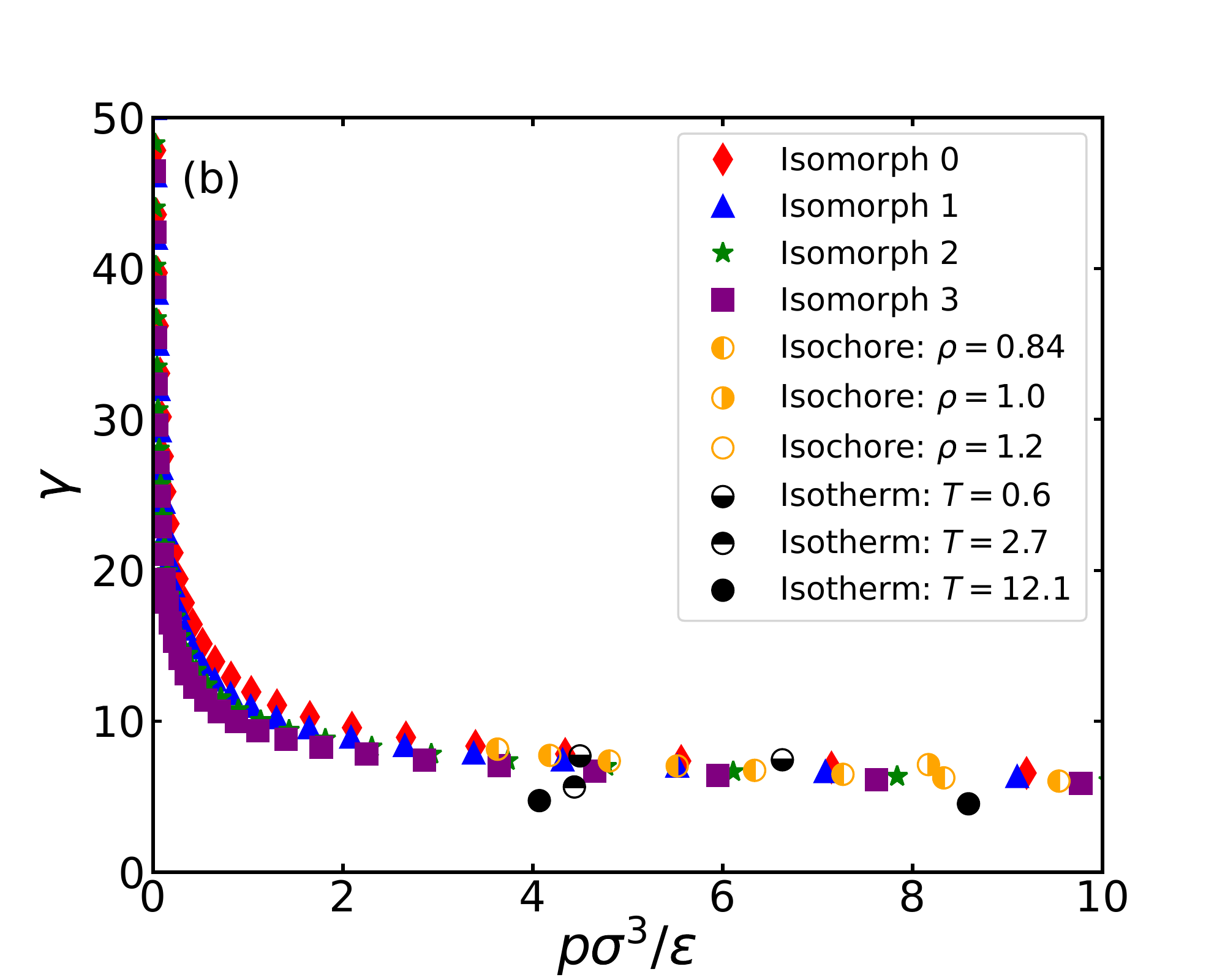}
	\includegraphics[width=5.3cm]{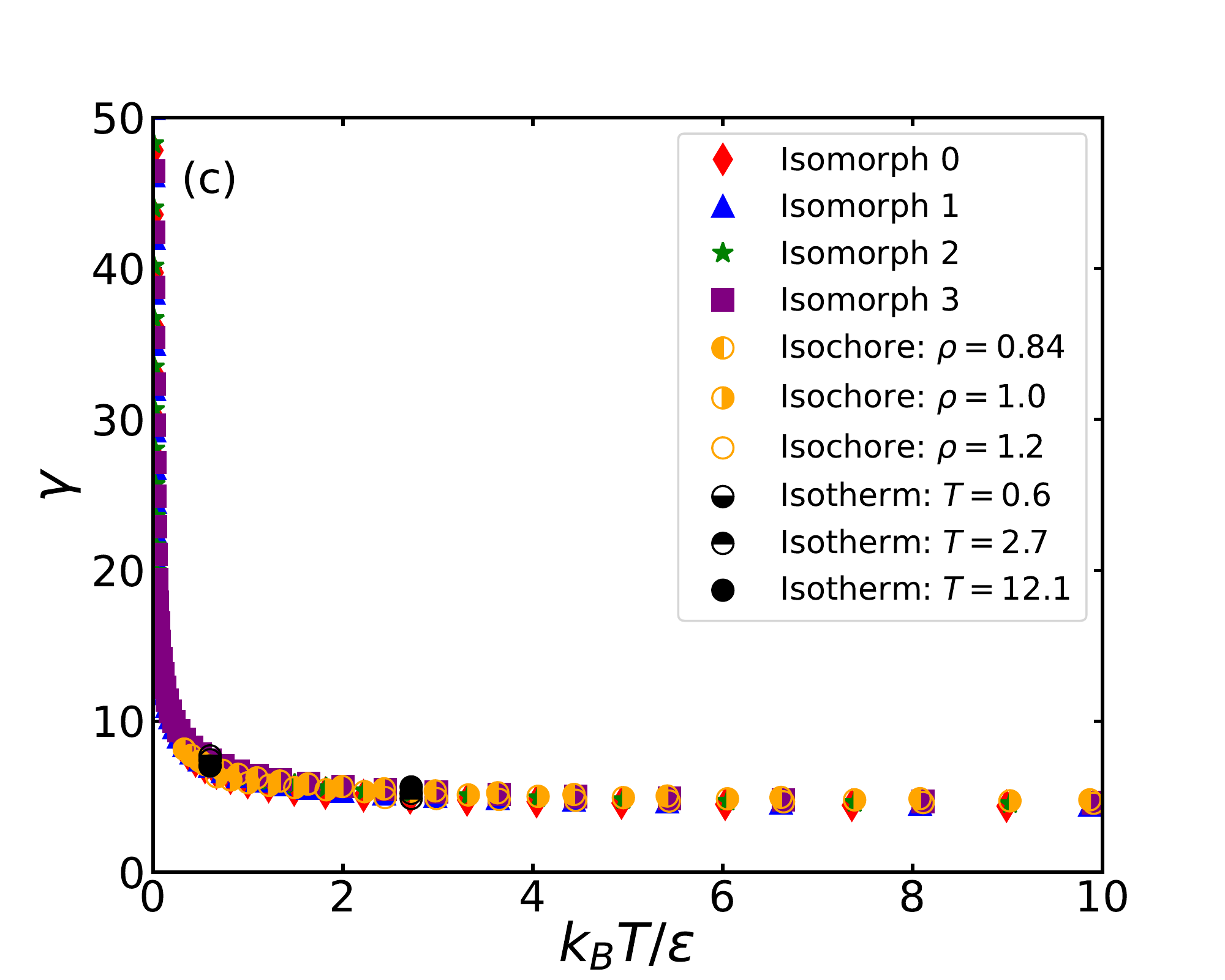}
	\includegraphics[width=5.3cm]{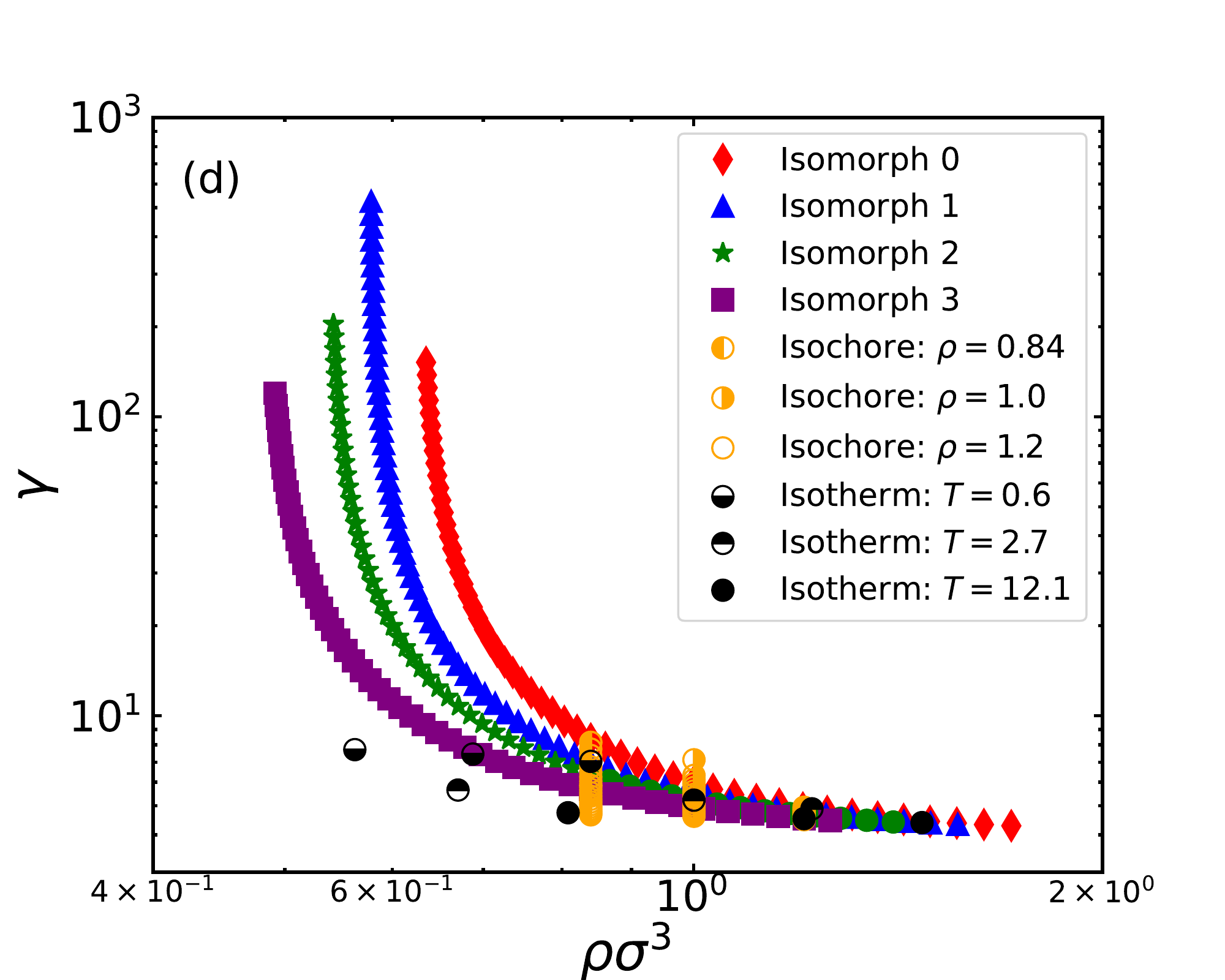}
	\includegraphics[width=5.3cm]{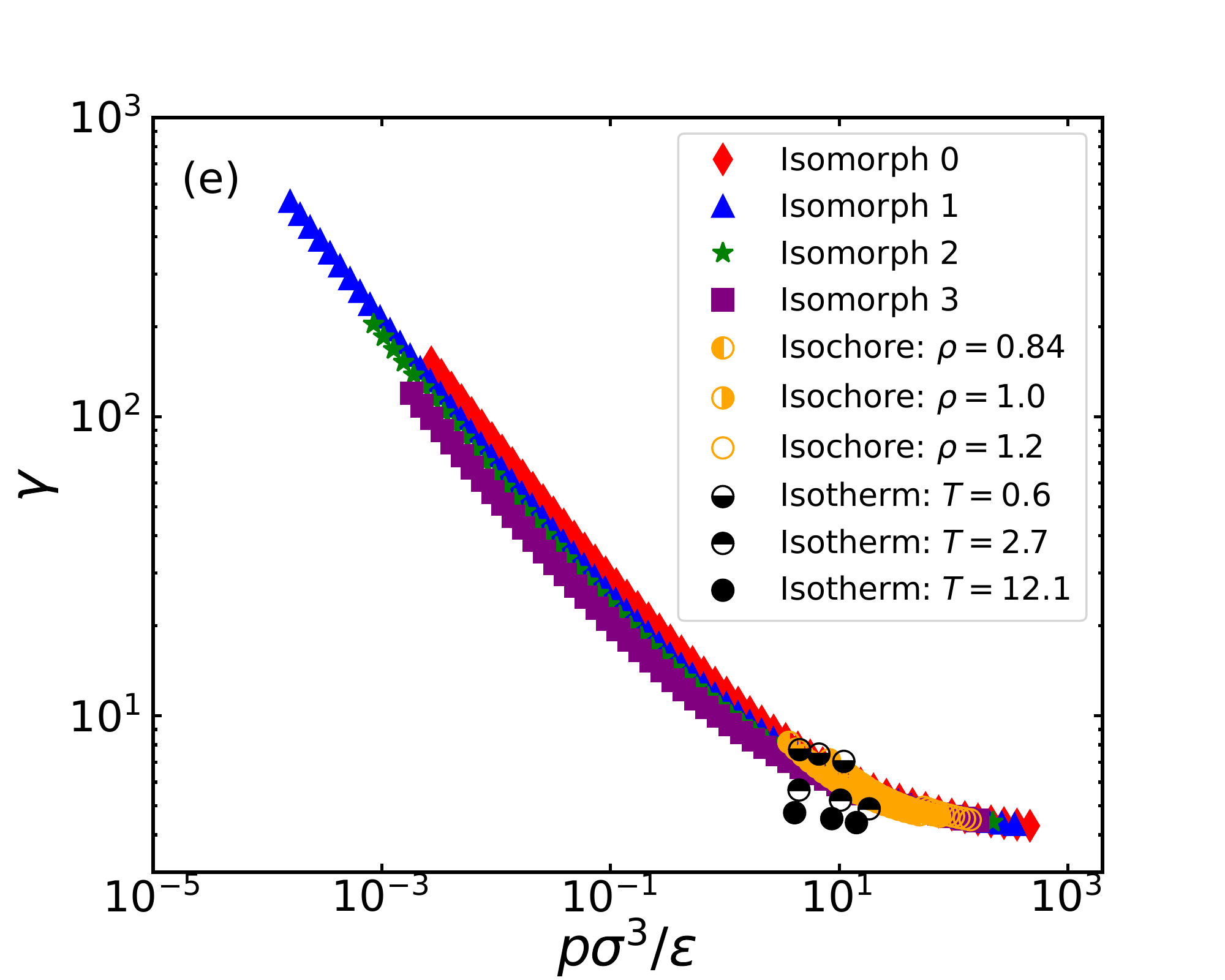}
	\includegraphics[width=5.3cm]{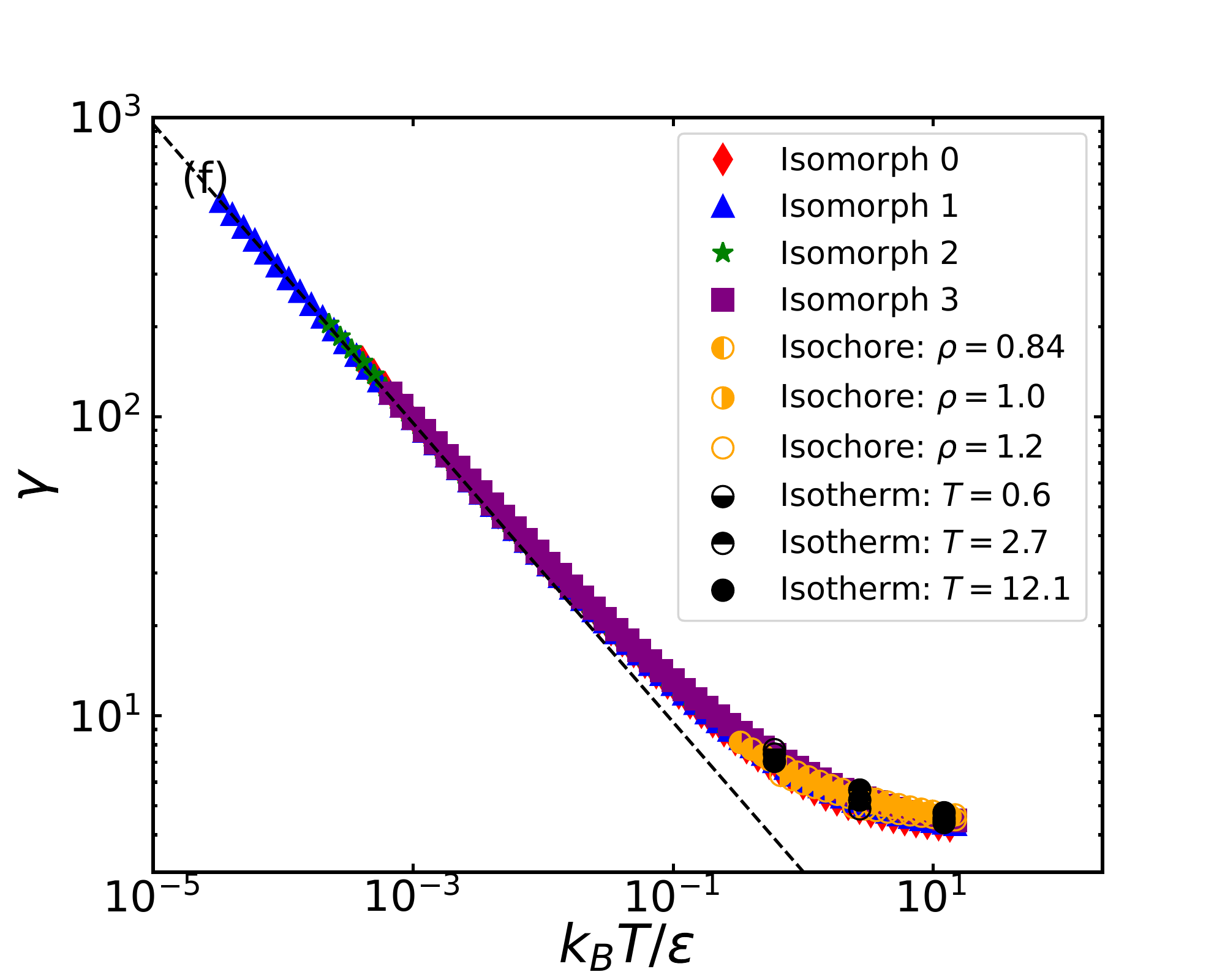}	
	\caption {\label{fig3} The density-scaling exponent $\gamma$ defined in \eq{gamma} for the state points studied (\fig{fig1}). Full symbols are isomorph state-point data, half open circles are isochore and isotherm data. The top row gives data for state points with $\gamma$ below 50, the bottom row gives data for all state points.
		(a) $\gamma$ as a function of the density;
		(b) $\gamma$ as a function of the pressure; 	
		(c) $\gamma$ as a function of the temperature;
		(d) $\gamma$ as a function of the density in a log-log plot;
		(e) $\gamma$ as a function of the pressure in a log-log plot;
		(f) $\gamma$ as a function of the temperature in a log-log plot. 
		We see that $\gamma$ is primarily a function of the temperature. The dashed line in (f) marks the $T\to 0$ limit of the mean-field theory (\eq{prediction}).
}
\end{figure}

\Fig{fig3} gives data for the density-scaling exponent $\gamma$ at the state points simulated plotted in different ways, using the same symbols as in \fig{fig2}. We see that $\gamma$ increases monotonically as either density, pressure, or temperature is lowered, eventually reaching values above 500. Figure 3(a) shows $\gamma$ as a function of the density $\rho$. Clearly, knowledge of $\rho$ is not enough to determine $\gamma$, implying that \eq{hgamma} does not apply for the WCA system. It has been suggested that $\gamma$ is controlled by the pressure \cite{cas20}. This works better than the density for collapsing data, but there is still some scatter ((b)). \Fig{fig3}(c) plots $\gamma$ as a function of the temperature. We here observe a quite good collapse, concluding thus $\gamma$ is primarily controlled by the temperature. \Fig{fig3}(d), (e), and (f) show data for all the state points simulated in a logarithmic plot as functions of density, pressure, and temperature, respectively.

\section{Mean-field theory for $R$ and $\gamma$ at low densities}\label{Sec4a}

This section presents a mean-field theory for estimating the virial potential-energy correlation coefficient $R$ and the density-scaling exponent $\gamma$. Along the lines of Refs. \onlinecite{EXPI,EXPII,mai16,mai20}, we assume that the individual pair interactions are statistically independent; this is expected to be a good approximation at relatively low densities. 

In MD units the truncated WCA pair potential \eq{wca} is 

\begin{equation}
v(r)=4r^{-12}-4r^{-6}+1 \textrm{ for } r<r_c\equiv 2^{1/6}=1.122\ldots
\end{equation}
and zero otherwise. The virial of the configuration $\bR$ is given by $W(\bR) = \sum_{i>j}^N w(r_{ij})$ in which the pair virial is defined as $w(r)\equiv -(r/3)v'(r)$ \cite{tildesley}. Although the WCA potential is our primary focus, the arguments given below apply to any truncated purely repulsive potential. 

The general the partition function of the configurational degrees of freedom is given by $Z\propto \int_{V^N} d\br_1...d\br_N\exp(-\sum_{i<j}v(r_{ij})/k_BT)$ in which $r_{ij}=|\br_i-\br_j|$. At low densities it is reasonable to regard the pair distances as uncorrelated, i.e., to treat the interactions in a mean-field way. This leads to the approximation $Z\propto Z_s^N$ in which $Z_s=\int_V d\br\exp(-v_s(\br)/k_BT)$ is the partition function of a single particle moving in the potential $v_s(\br)$ of all other particles frozen in space. In the low-density limit, none of the frozen particles ``overlap'' and $Z_s$ has consequently two contributions, one for the positions for which $v(\br)=0$ and one for the positions at which the particle interacts with one of the frozen particles. The former is the ``free'' volume that in the low-density limit is approaches the entire volume $V$. The latter is $N$ times the following integral (putting for simplicity $k_B=1$ in this section),

\begin{equation}
Z_1(T) =  \int_0^{r_c} 4\pi r^2 \exp(-v(r)/T)dr\,.
\end{equation}
In terms of $Z_1(T)$ the single-particle partition function is thus in the thermodynamic limit given by

\be
Z_s(\rho,T)/N=Z_1(T)+1/\rho\,.
\ee

\begin{figure}
	\centering
	\includegraphics[width=6cm]{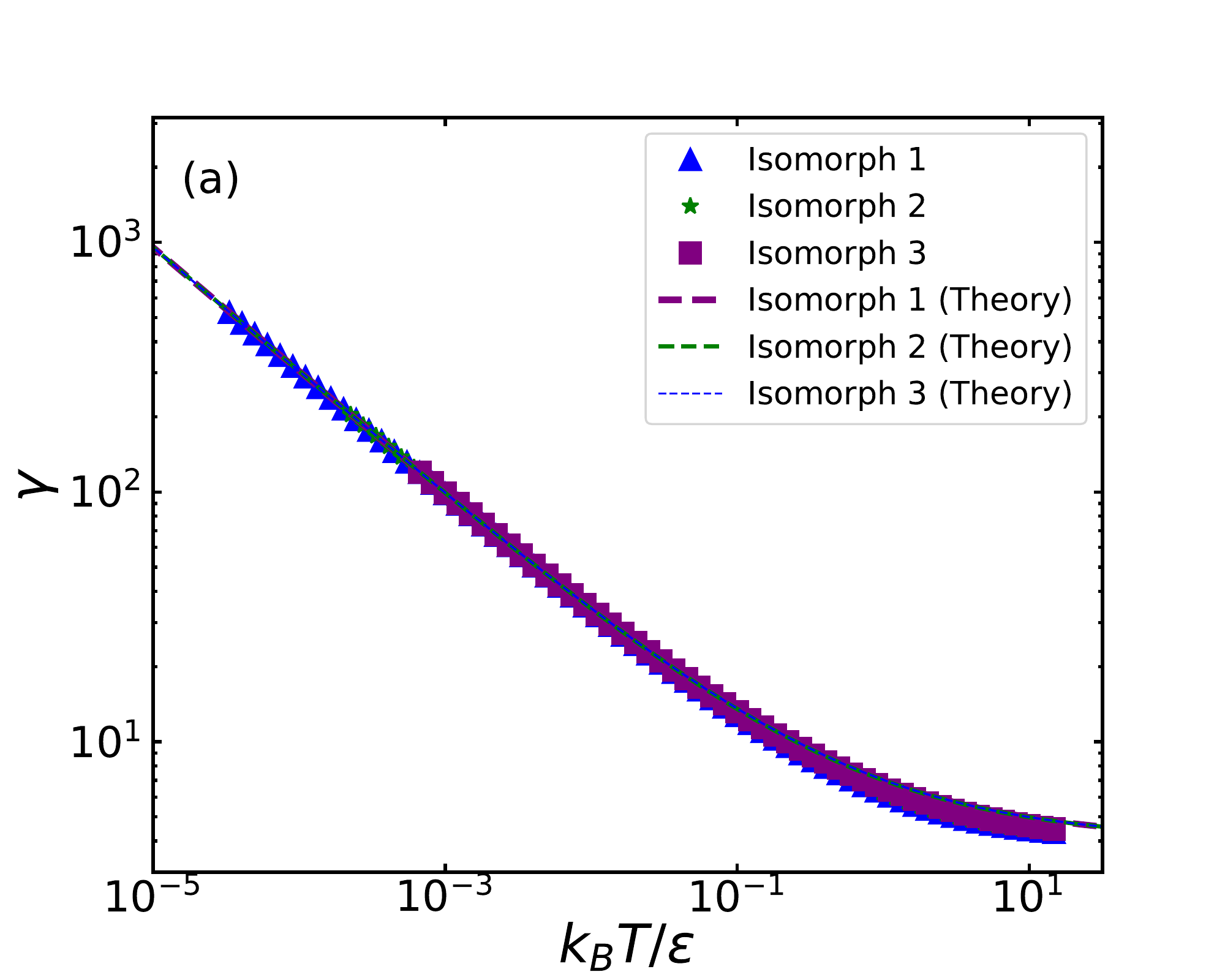}
	\includegraphics[width=6cm]{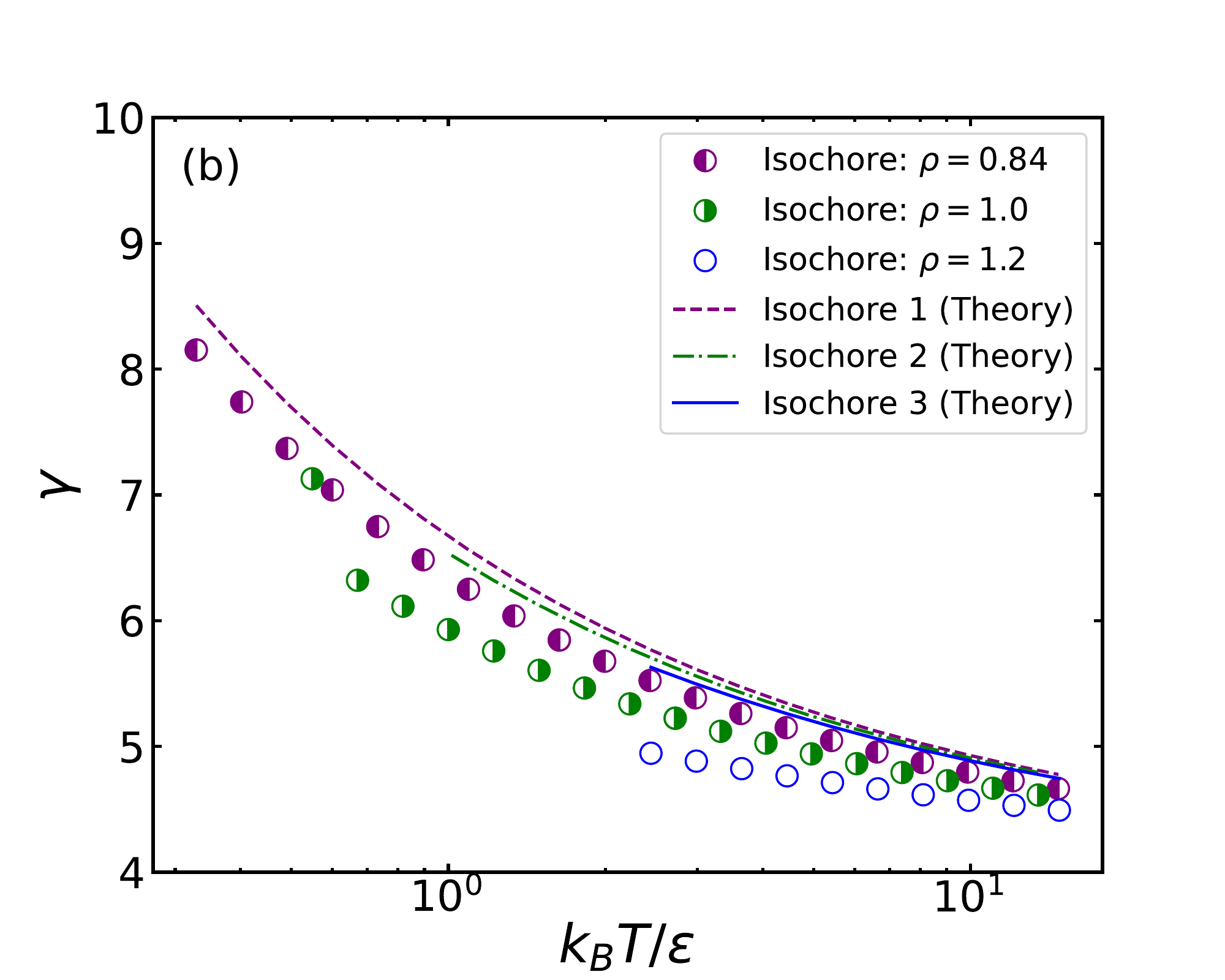}
	\includegraphics[width=6cm]{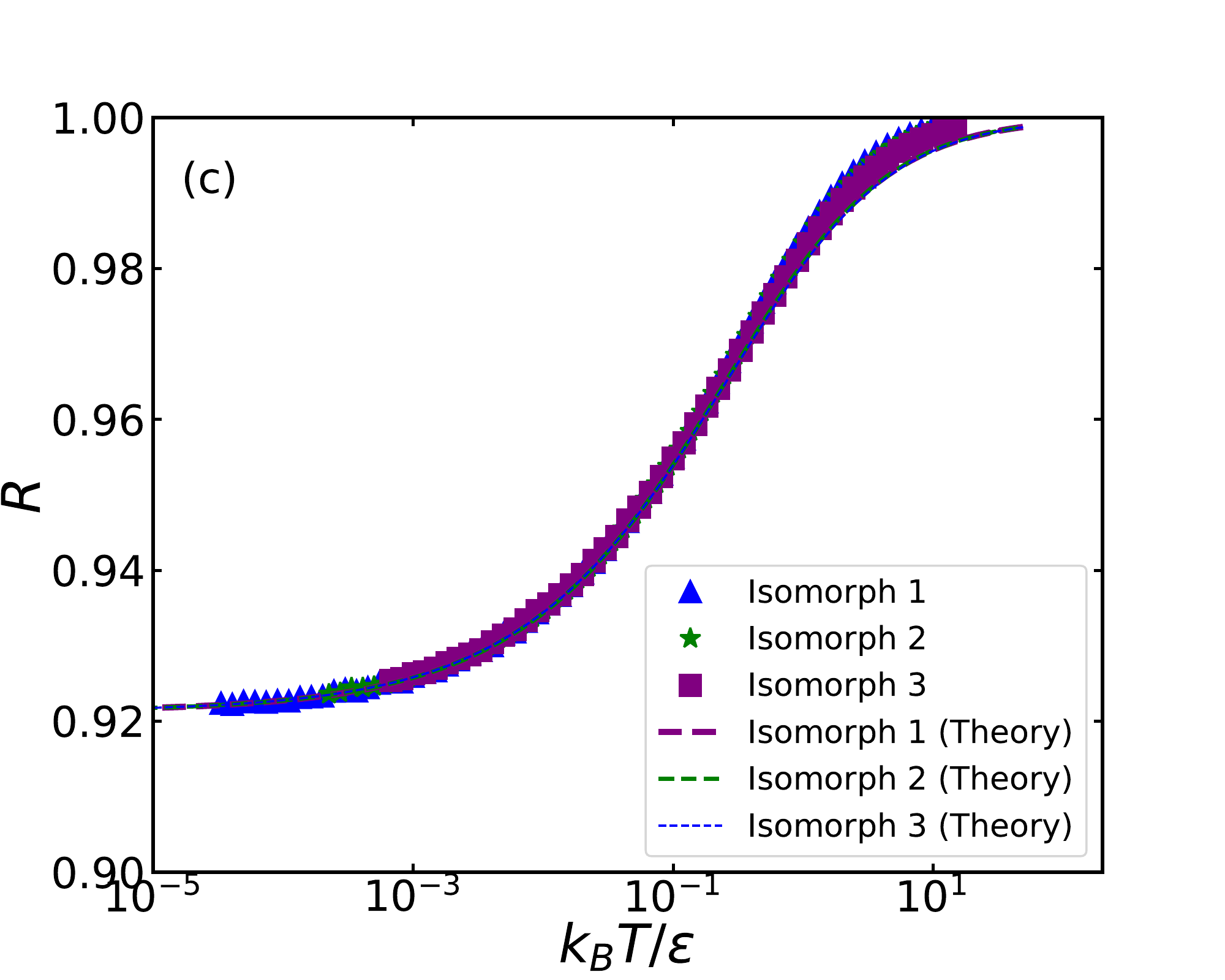}
	\includegraphics[width=6cm]{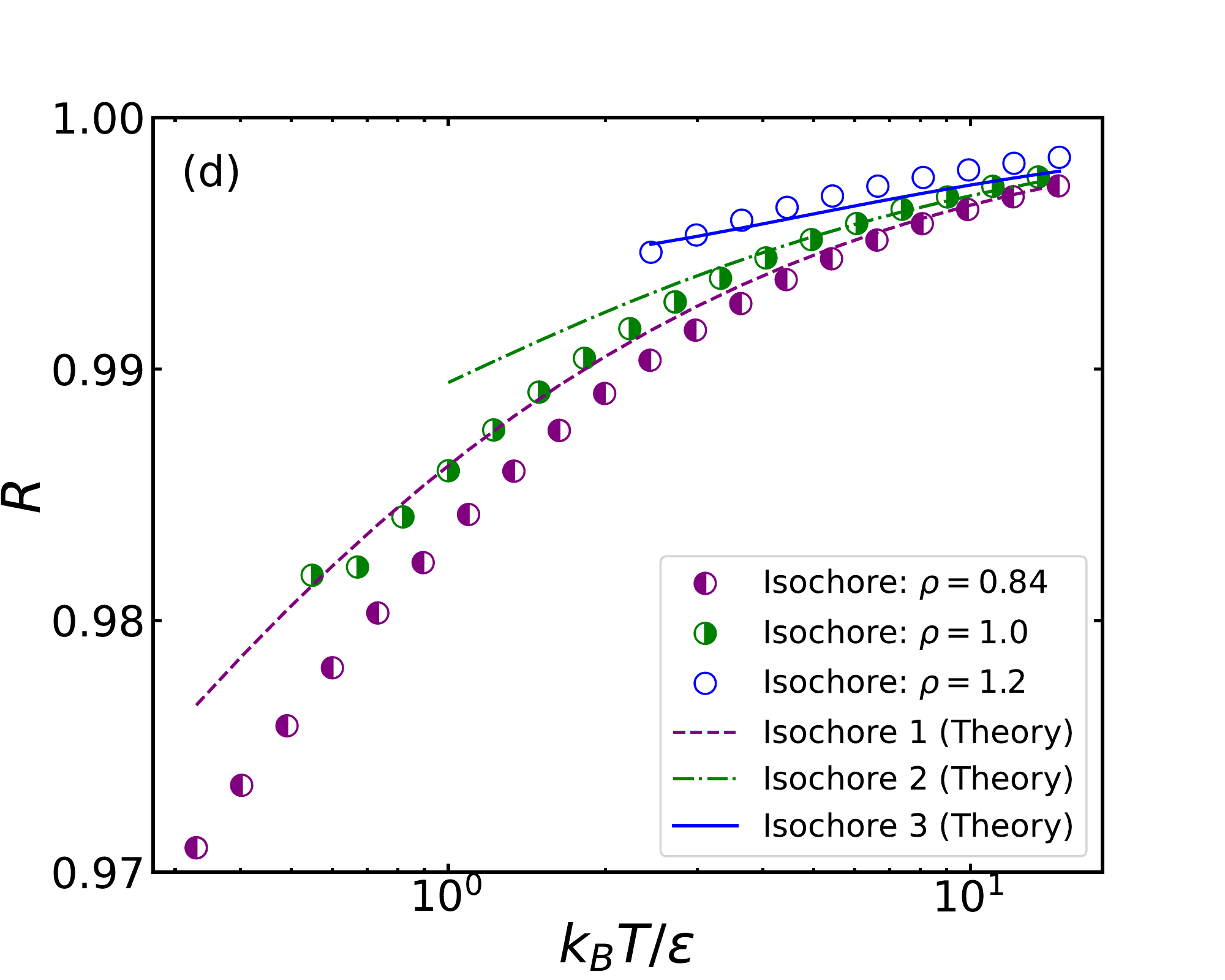}	
	\caption{\label{fig5}Comparing the predictions of the mean-field theory for $\gamma$ and $R$ as functions of the temperature (dashed lines) to simulation results. (a) and (c) show results along the three isomorphs. (b) and (d) show results along the three isochores, focusing on higher densities where the mean-field theory is not expected to be accurate.}
\end{figure}

Based on the above, any pair-defined quantity $A(r)$ that is zero for $r>r_c$ has an expectation value that is computed as (in which $p(r)=4\pi r^2 \exp(-v(r)/T)$ is the unnormalized probability)

\begin{equation}\label{eq:expectation_value}
\langle A \rangle = \int_{0}^{r_c}A(r)p(r)dr/Z_s(\rho,T)\,.
\end{equation}
Based on \eq{gamma} and \eq{R_def} one gets

\begin{equation}
\gamma(\rho, T) = \frac{\langle wv\rangle-\langle w\rangle\langle v\rangle}{\langle v^2\rangle-\langle v\rangle^2}
\end{equation}
and 

\begin{equation}
R(\rho, T) = \frac{\langle wv\rangle-\langle w\rangle\langle v\rangle}{\sqrt{(\langle w^2\rangle-\langle w\rangle^2)(\langle v^2\rangle-\langle v\rangle^2)}}.
\end{equation}
\Fig{fig5} compares the predictions of the mean-field theory to data along isomorphs and isochores. There is good overall agreement. Systematic deviations are visible in (b) and (d), however, which focus on densities that are now low enough to avoid frozen-particle overlap.

We proceed to discuss the low-density limit in which $Z_s\to\infty$. Terms that involve a single expectation value ($\langle v^2\rangle$, $\langle w^2\rangle$, and $\langle wu\rangle$) scale as $1/Z_s$ while terms that involve a multiplication of expectation values, i.e., $\langle v\rangle^2$, $\langle w\rangle^2$, and $\langle v\rangle\langle w\rangle$, scale as $1/Z_s^2$. Consequently, at low densities one can neglect terms that involve multiplications of expectation values \cite{mai16,EXPI,EXPII,mai20}, leading to

\begin{equation}
\gamma(T) = \langle wv\rangle/\langle v^2\rangle \textrm{ for } \rho\rightarrow 0
\end{equation}
and 

\begin{equation}
R(T) = \langle wv\rangle/\sqrt{\langle w^2\rangle\langle v^2\rangle} \textrm{ for } \rho\rightarrow 0.
\end{equation}
Note that these averages do not depend on $Z_s$ since both numerators and denominators scale as $1/Z_s$. This implies that $\gamma$ and $R$ at low densities depend only of $T$, which explains the observation in \fig{fig3}. 

Consider now the further limit of low temperature. In that case the probability distribution $p(r)$ concentrates near $r_c$ and one can expand around $x\equiv r_c-r=0$ by writing the pair potential as

\begin{equation}
v(x) = k_1x+ k_2x^2/2+k_3x^3/6+.....
\end{equation}
The pair virial then becomes \cite{II}

\begin{equation}
w(x) = (r_c-x)(k_1/3 + k_2x/3)\,+\,k_3r_cx^2/6+O(x^3)\,.
\end{equation}
For the WCA potential $k_1=0$ and $k_2=36\sqrt[3]{4}$. Since $p(x)$ is concentrated near $x=0$, the upper limit of the integral \eq{eq:gaussian} may be extended to infinity, leading to

\begin{equation}\label{eq:gaussian}
\langle A\rangle =  \int_0^{\infty} A(x) p(x)dx/Z \,\,\, (T \rightarrow 0)
\end{equation}
in which

\begin{equation}
p(x) = 4\pi (r_c-x)^2 \exp(-k_2x^2/(2T))\,.
\end{equation}
The Gaussian integrals can be evaluated by hand or, e.g., using the SymPy Python library for symbolic mathematics. We find that $\gamma$ and $R$ in the low-density limit are given by

\begin{equation}\label{prediction}
\gamma_0 = \frac{4r_{c} \sqrt{2k_{2}} }{9 \sqrt{\pi T}} = \frac{16}{3 \sqrt{\pi T}}\,\,\, (T \rightarrow 0)
\end{equation}
and

\begin{equation}\label{R0}
R_0 = \sqrt{\frac{8}{3 \pi}} = 0.921\ldots\,\,\, (T \rightarrow 0)\,.
\end{equation}

\begin{figure}
	\centering
	\includegraphics[width=6cm]{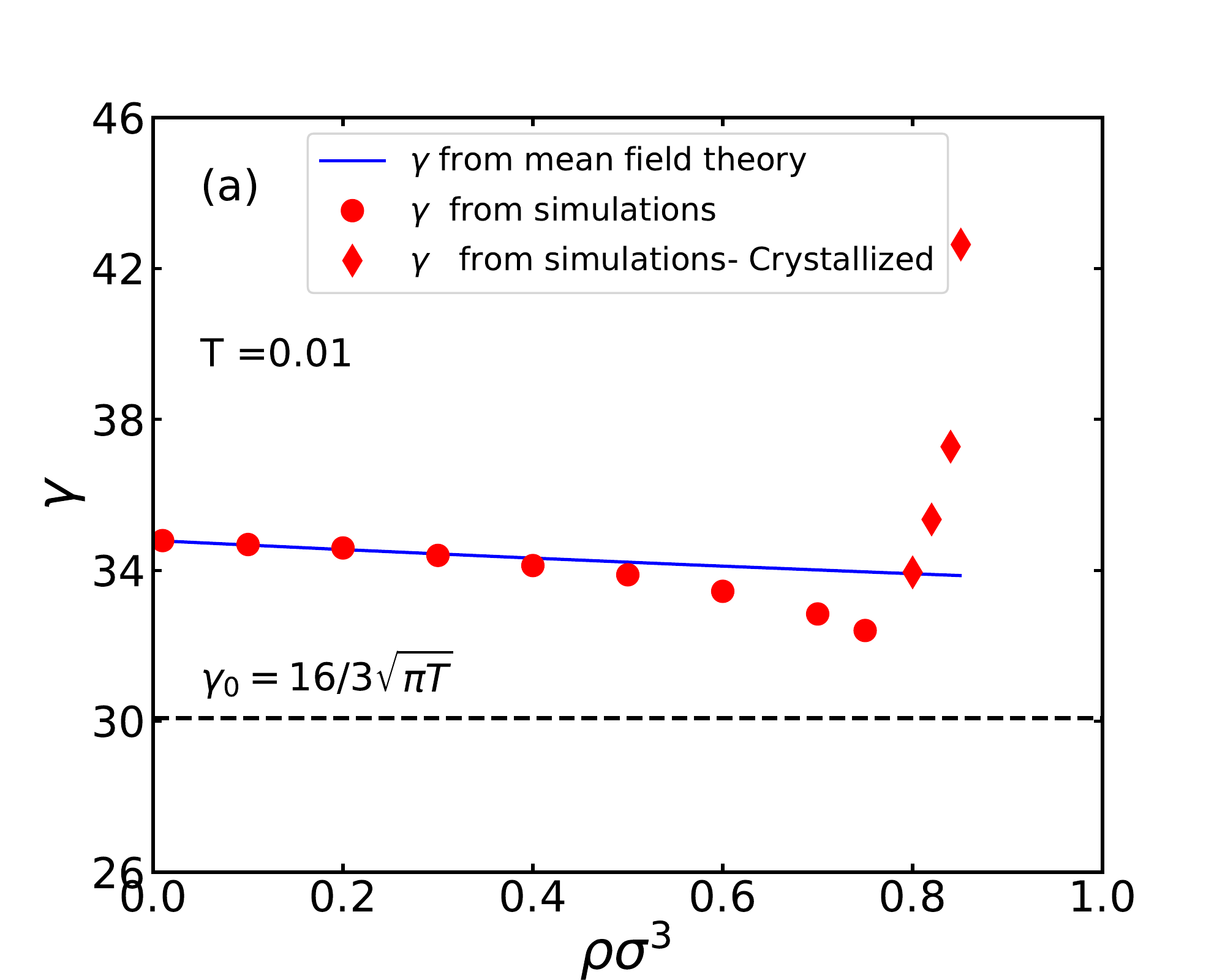}
	\includegraphics[width=6cm]{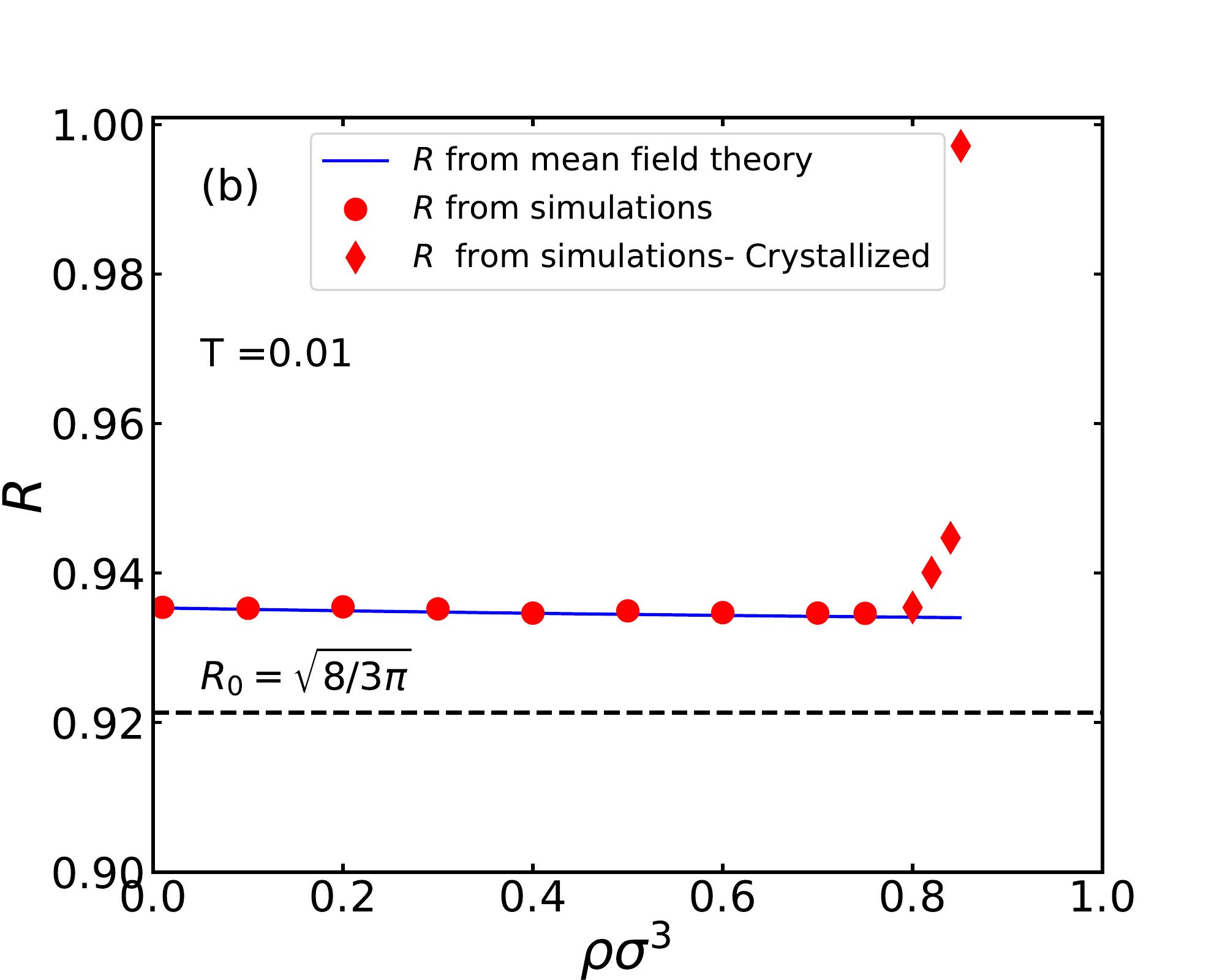}	
	\caption{\label{fig6}The density dependence of (a) $\gamma$ and (b) $R$ at $T=0.01$. Results are also shown for high-density samples that crystallized during the simulations.}
\end{figure}

\Fig{fig6} shows the mean-field predictions for $\gamma$ and $R$ at $T=0.01$ plotted as a function of the density. As expected, the theory works well at low densities, even though one is here still not quite at the $T\to 0$ limit marked by the horizontal lines.

\section{Variation of structure and dynamics along isotherms, isochores, and isomorphs}\label{Sec4}

\begin{figure}[h]
	\includegraphics[width=12cm]{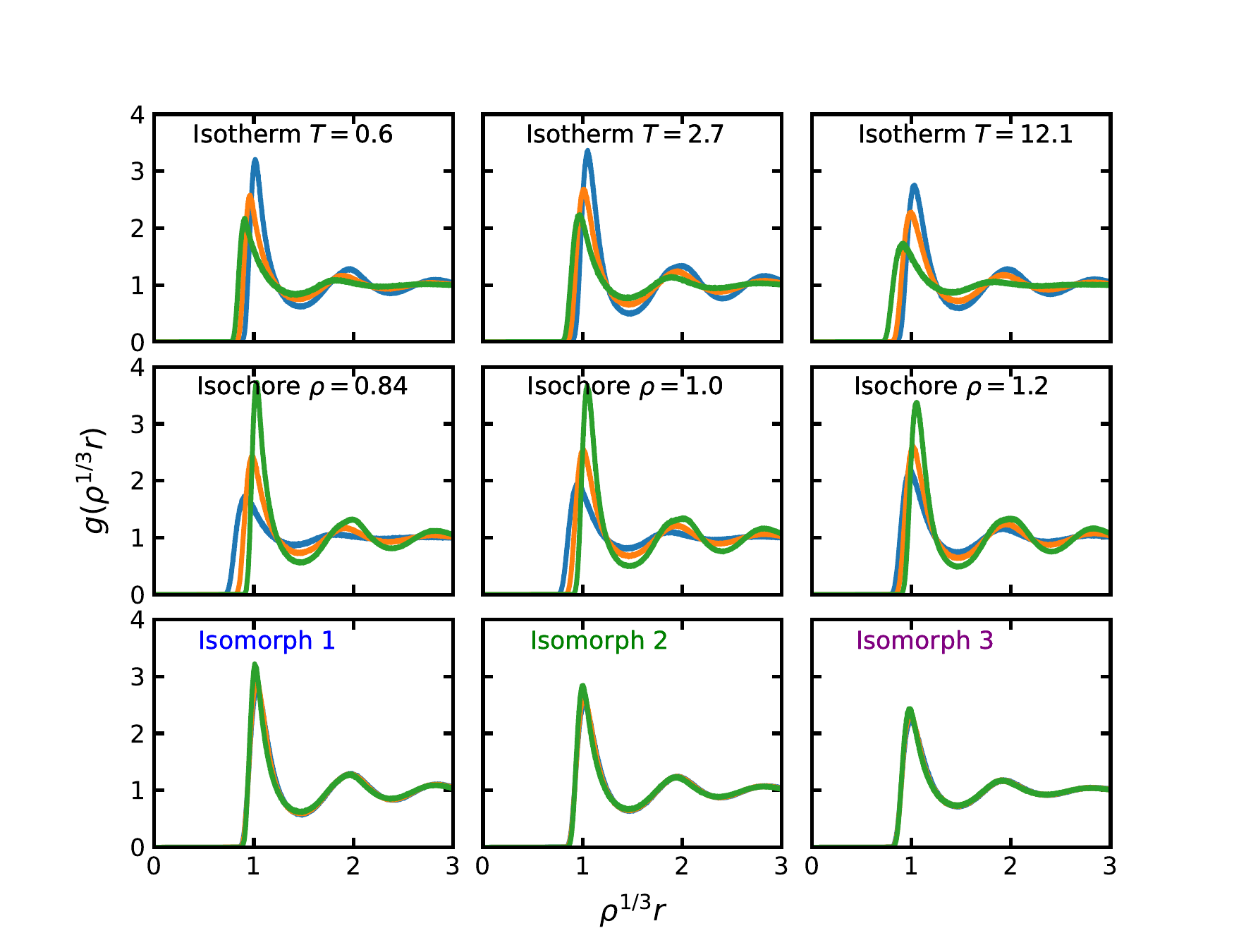}
	\begin{minipage}{1.1\textwidth}
	\caption {\label{fig7} Reduced-unit radial distribution functions (RDF) for the three isotherms, isochores, and isomorphs (\fig{fig1}). The green curves give the lowest temperature/density, the orange curves give the mid temperature/density, and the blue curves give the highest temperature/density. Although the first-peak maximum is not entirely isomorph invariant, in comparison to isotherms and isochores we see a good RDF invariance along the isomorphs. This is the case even though the density variation of the isotherms and the temperature variation of the isochores are somewhat smaller than those of the isomorphs (compare \fig{fig1}). 
	\textit{Isotherms:} The green curves give data for $(\rho,T)=$ (0.56, 0.60), (0.82, 2.72), (0.81, 12.1), 
	the orange curves for $(\rho,T)=$ (0.69, 0.60 ), (1.0, 2.72), (1.21, 12.1), 
	and the blue curves for  $(\rho,T)=$ (0.84, 0.60), (1.22, 2.72), (1.47, 12.1).	
	\textit{Isochores:} The green curves give data for $(\rho,T)=$ (0.84, 0.33), (1.00, 0.82), (1.21, 2.44), 
	the orange curves for  $(\rho,T)=$ (0.84, 1.99), (1.00, 3.32), (1.21, 6.64), 
	and the blue curves for  $(\rho,T)=$ (0.84, 14.72), (1.00, 13.46), (1.21, 14.78).
	\textit{Isomorphs:} The green curves give data for the reference state points $(\rho,T)=$ (0.84, 0.60), (0.84, 1.00), (0.84, 2.00), the orange curves for  $(\rho,T)=$ (1.06, 2.43), (1.04, 3.32), (0.94, 3.64), and the blue curves for  $(\rho,T)=$ (1.57, 14.72), (1.40, 13.46), (1.26, 14.78).	
}
\end{minipage}
\end{figure}

The considerable $\gamma$ variation of the WCA system means that it cannot be described approximately by an Euler-homogeneous potential-energy function. This section investigates to which degree the reduced-unit structure and dynamics are, nevertheless, invariant along isomorphs 1-3. Isomorph invariance is rarely exact, so in order to put the simulation results into perspective, we present also results for the variation of the reduced-unit structure and dynamics along isotherms and isochores (\fig{fig1}). As a measure of the structure, we look at the reduced radial distribution function (RDF) as a function of the reduced radial distance. As a measure of the dynamics, we look at the reduced mean-squared displacement (MSD) as a function of the reduced time, as well as on the reduced diffusion coefficient $\tilde{D}$ identified from the long-time MSD.

Starting with structure, \fig{fig7} shows reduced-unit RDF data along the three isotherms, isochores, and isomorphs. The isotherms span almost the same density range and the isochores span almost the same temperature range as the corresponding isomorphs (restricted to the equilibrium liquid phase, i.e., to data above the freezing line). Along the isomorphs the RDFs show some variation at the first peak maximum (lowest row), but in comparison to the isotherms and isochores, there is excellent overall isomorph invariance of the RDF. 

For all three isomorphs we find that the peak height increases as the temperature decreases. This is an effect of larger $\gamma$ resulting in a higher first peak, which may be understood as follows. Consider the IPL pair-potential system with $v(r)\propto r^{-n}$, which has $\gamma=n/3$ and perfect isomorphs \cite{hey15}. The larger $n$ is, the more harshly repulsive are the forces. From the Boltzmann probability of finding two particles at the distance $r$, $\propto\exp(-v(r)/k_BT)$, it follows that particle near encounters become less likely as $n\to\infty$, thus suppressing the RDF at distances below the first peak. If there is isomorph invariance of the number of particles within the first coordination shell, as $n$ increases some of the RDF must therefore move from small $r$ to larger $r$ within the first coordination shell, resulting in a higher first peak. This argument has recently been confirmed by the observation that the bridge function, a fundamental quantity of liquid-state theory \cite{han13}, is isomorph invariant to a very good approximation \cite{cas21}. A similar increase of the height of the first RDF peak with increasing $\gamma$ has been observed for the EXP system (Fig. 5 in Ref. \onlinecite{EXPII}). In that case it was a much less dramatic effect, however, because the EXP system's $\gamma$ variation at the investigated state points covered less than a factor of 3 compared to more than a factor of 100 for the WCA state points studied here. Interestingly, for both systems the data imply that $\gamma\to\infty$ as $T\to 0$ along an isomorph, i.e., both systems become more and more hard-sphere like as the temperature is lowered.

\begin{figure}[h]
	\includegraphics[width=12cm]{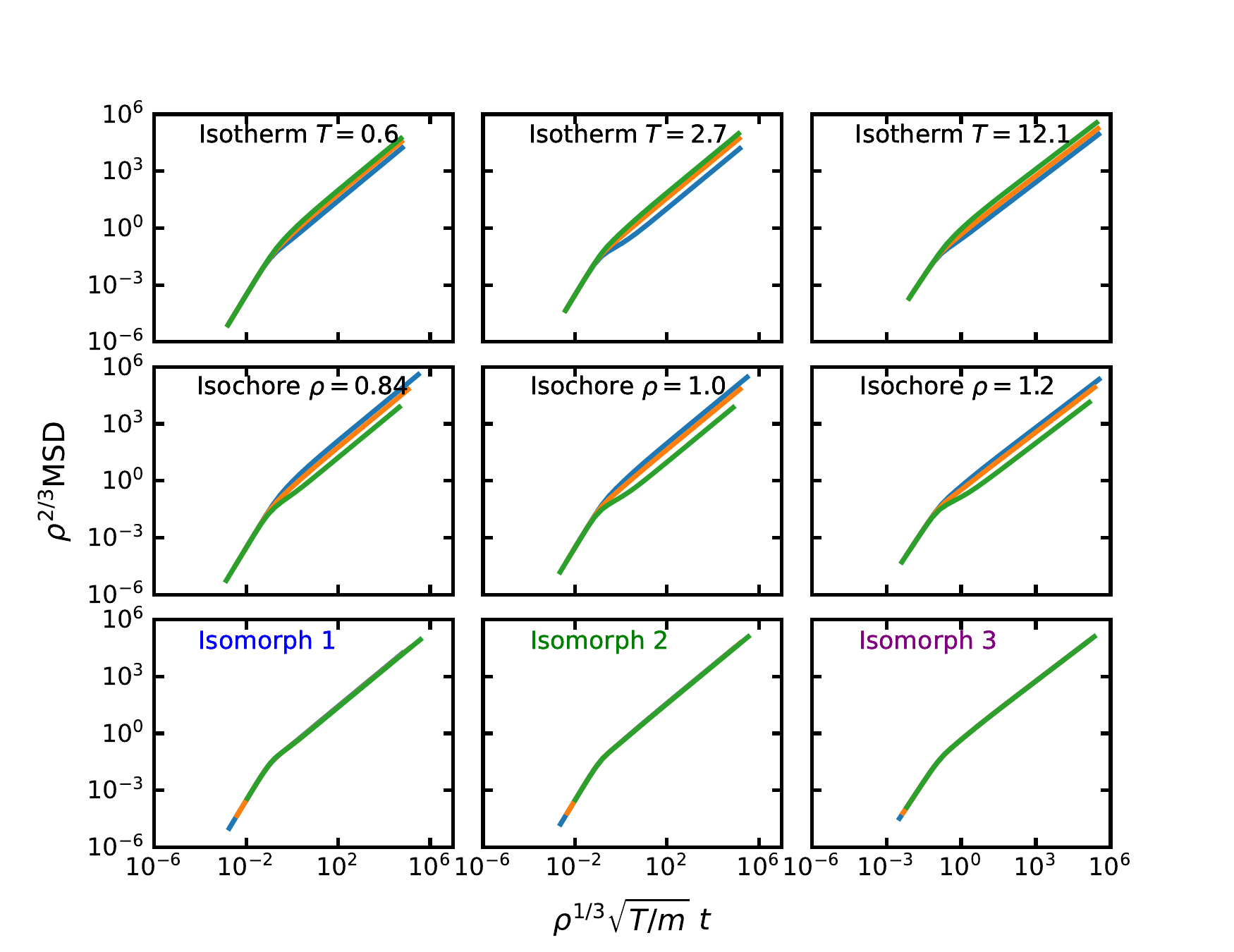}
	\caption{\label{fig8} Reduced-unit radial mean-squared displacement (MSD) plotted against time for the three isotherms, isochores, and isomorphs (\fig{fig1}). The state points and color codings are the same as in \fig{fig7}. The dynamics is isomorph invariant to a very good approximation.}
\end{figure}

Proceeding to investigate the dynamics, \fig{fig8} shows data for the reduced-unit MSD as a function of the reduced time along the three isotherms, isochores, and isomorphs. There is only invariance along the isomorphs. Along the isotherms, the lowest density (green) give rise to the largest reduced diffusion coefficient. This is because the mean collision length increases when density is decreased. Along the isochores, the lowest temperature (green) has the smallest reduced diffusion coefficient. This is because the effective hard-sphere radius increases when temperature is decreased, leading to a smaller mean-collision length. In MD units, the MSDs are also not invariant along the isotherms or isochores (data not shown); thus the lack of invariance for the isotherms and isochores is not a consequence of the use of reduced units. In regard to the isomorph data, with \fig{fig7} in mind we conclude that the non-invariant first-peak heights of the RDFs along the isomorphs has little influence on the dynamics. This is consistent with expectations from liquid-state quasiuniversality, according to which many systems have structure and dynamics similar to those of the EXP generic liquid system, which as mentioned also exhibits varying first-peak heights along its isomorphs \cite{EXPII}.

\begin{figure}[h]
	\includegraphics[width=9cm]{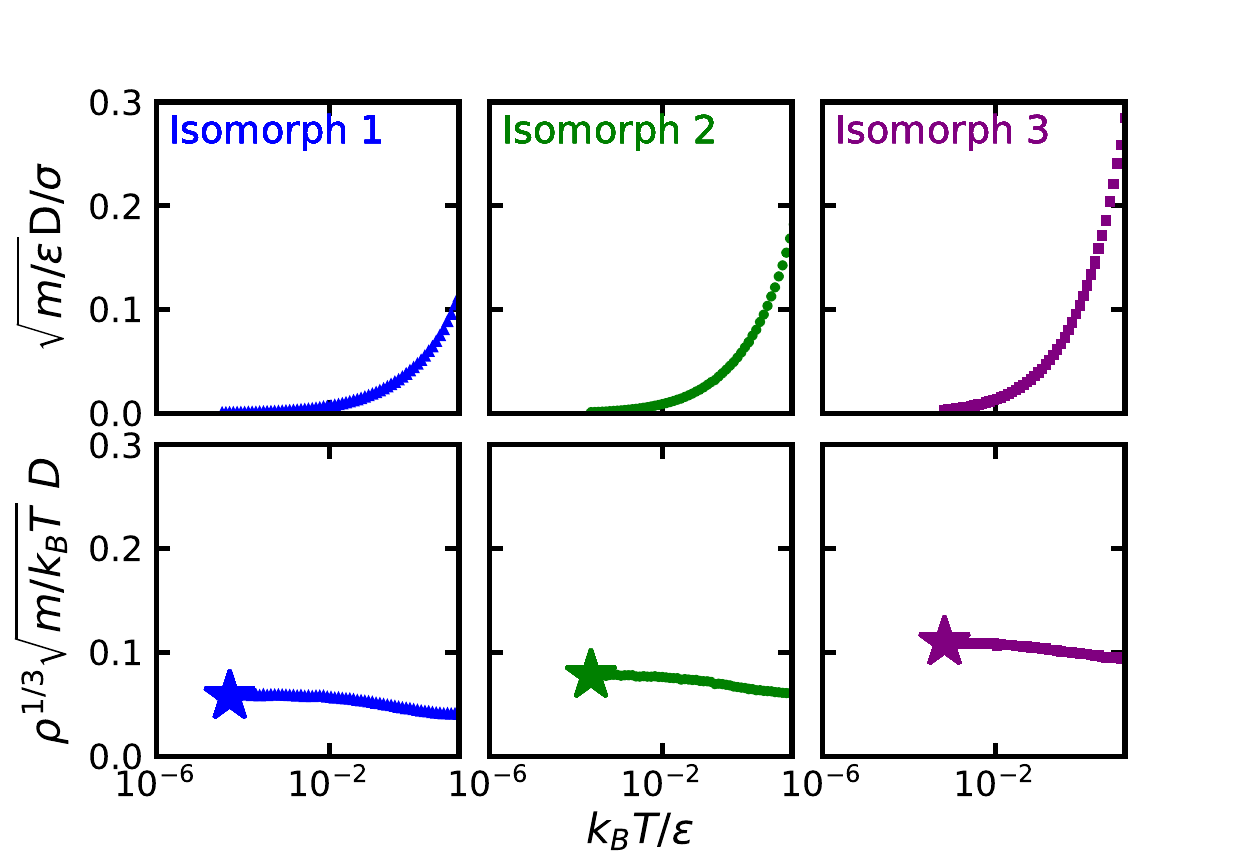}
	\caption{\label{fig9} Diffusion coefficients along isomorphs 1-3 in MD units (upper row) and in reduced units (lower row), plotted as functions of the logarithm of the temperature. When given in MD units, the diffusion coefficients vary significantly along the isomorphs, while they are fairly constant in reduced units. This illustrates the importance of using reduced units when checking for isomorph invariance. From end point to end point of the isomorphs, the variation in the reduced diffusion coefficient $\tilde D$ is, respectively, 39\%, 23\%, and 14\%. The corresponding numbers are 1000\%, 880\%, and 549\% along the isochores, and 214\%, 893\%, and 305\% along the isotherms.}
\end{figure}

The reduced diffusion coefficient $\tilde{D}\equiv\rho^{1/3}\sqrt{m/k_BT}\,D$ is extracted from the data in \fig{fig8} by making use of the fact that the long-time reduced MSD is $6\tilde{D}\tilde{t}$. \Fig{fig9} shows how both $D$ and $\tilde{D}$ vary along the three isomorphs. The upper figures demonstrate a large variation in $D$ along each isomorph. The lower figures show $\tilde{D}$, which is rigorously invariant for a system with perfect isomorphs ($R=1$). This is not the case for the WCA system, but the variation is below 40\% for all three isomorphs in situations where the temperature varies by more than four orders of magnitude. Thus the reduced diffusion coefficient is isomorph invariant to a good approximation.

\begin{figure}[h]
	\includegraphics[width=6cm]{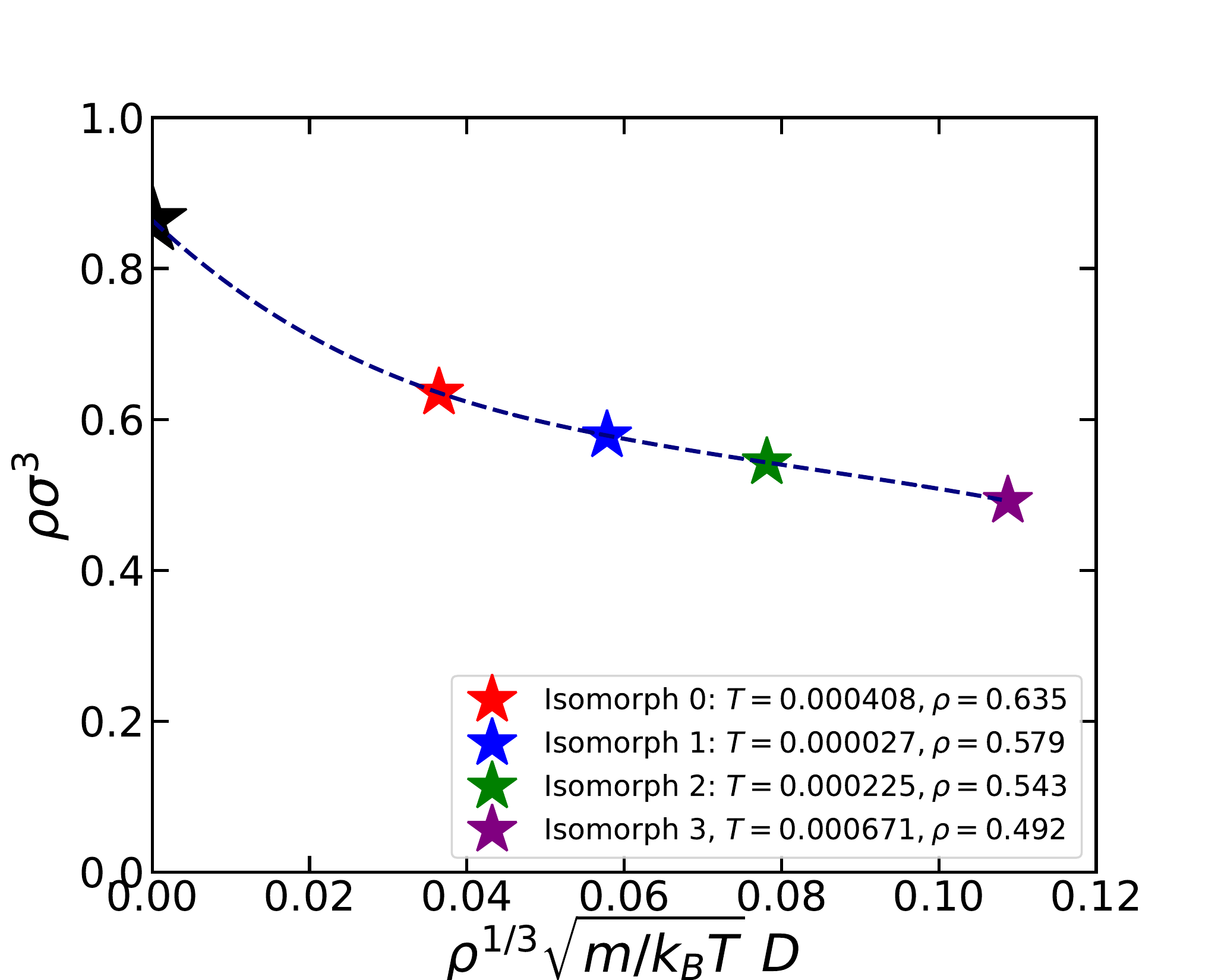}
	\caption{\label{fig13} The reduced diffusion coefficients at the lowest temperature and density for isomorphs 1-3 supplemented by data for isomorph 0, plotted versus the density of the lowest-temperature state point simulated on the isomorph in question. The points are fitted by a cubic spline function (dashed curve), which by construction goes through the random close-packing (rcp) density ($\rho=0.864$) marked by the black star on the y-axis. As rcp is approached, one expects $\tilde{D}\to 0$ because the system jams. This is consistent with our data. The rcp density is calculated as follows. With $r_c = 2^{1/6} $ one finds $V_{\rm sphere} = \pi r_c^3/6 =  0.74048$. The rcp volume fraction is roughly 64\%; putting this equal to $\rho V_{\rm sphere}$, one arrives at $\rho=0.864$.}
\end{figure}

\Fig{fig9} suggests that $\tilde{D}$ stabilizes as $T\to 0$, and for each isomorph one can tentatively identify this low-temperature limit. \Fig{fig13} plots estimates of these limiting values obtained at the lowest density simulated on each isomorph. An obvious question is: which density corresponds to $\tilde{D}=0$? At very low temperatures, because $\gamma$ becomes very large the WCA system behaves increasingly as a system of hard spheres (HS). The disordered HS system has a maximum density corresponding to the random closed-packed (rcp) structure at roughly 64\% packing fraction. In \fig{fig13}, the black star at $\tilde{D}=0$ marks the corresponding density. Our data are consistent with a convergence to this point.

\section{Discussion}\label{Sec5}

We have studied three isomorphs of the WCA system and showed that along them the density-scaling exponents vary by more than a factor of 100. This extreme variation means that the WCA system can not be considered as an effective IPL system \cite{II}. In the LJ case, the pair potential may be approximated by the so-called extended IPL (eIPL) pair potential, which is a sum of an IPL term $\sim r^{-18}$, a constant, and a term proportional to $r$ \cite{II}. The latter two terms contribute little to the fluctuations of virial and potential energy \cite{II}, which explains the strong correlations of the LJ system and why $\gamma$ is close to 6 (not to 4 as one might guess from the repulsive $r^{-12}$ term of the potential). The WCA situation is very different. Because the WCA system is purely repulsive, it has no liquid-gas phase transition and no liquid-gas coexistence region. This means that isomorphs may be studied over several orders of magnitude of temperature and, in particular, followed to very low temperatures. Interestingly, even here the strong-correlation property is maintained. At the same time, $\gamma$ increases in an unprecedented fashion. Despite this, the reduced-unit structure and dynamics are both invariant to a good approximation along the isomorphs. The significant difference between the LJ and WCA systems in regard to isomorph properties is also emphasized by the fact that the density-scaling exponent $\gamma$ of the LJ system is primarily a function of the density and well described by \eq{h}. This is explained by the above-mentioned approximate eIPL pair-potential argument \cite{II}. 

The finding that $R$ and $\gamma$ of the WCA system are both primarily functions of the temperature is accounted for by a mean-field theory based on the assumption of statistically independent pair interactions. The same feature is observed for the EXP pair-potential system \cite{EXPII}, and also here do both $R$ and $\gamma$ at low densities primarily depend on the temperature. Another situation where this is expected to apply is the repulsive Yukawa pair-potential system at low densities \cite{vel15,tol19}.

In summary, the WCA systems presents a striking case where the density-scaling exponent is very far from being constant throughout the thermodynamic phase diagram \cite{san19,cas19}. Nevertheless, the system R-simple and has good isomorph invariance of the structure and dynamics.

\begin{acknowledgments}
	This work was supported by the VILLUM Foundation's \textit{Matter} grant (16515).
\end{acknowledgments}

\section*{Appendix I: Using the Runge-Kutta method for tracing out isomorphs efficiently}

The density-scaling exponent $\gamma$ is the slope of the lines of constant $\Sex$ in the $(\ln T,\ln\rho)$ plane (\eq{gamma}). By numerical integration one can from \eq{gamma} compute the lines of constant $\Sex$, the configurational adiabats, which are isomorphs for any R-simple system. The density-scaling exponents required for the integration are determined from the thermal equilibrium virial potential-energy fluctuations in an $NVT$ simulation (\eq{gamma}). In the following we denote the \emph{theoretical slope} by $f$, i.e., the slope without the unavoidable statistical noise of any MD simulation. Let $(x,y)$ be $(\ln\rho,\ln T)$ (occasionally it is better to choose instead $(x,y)=(\ln T, \ln\rho)$). In this notation, let
 
\be
\frac{dy}{dx}= f(x,y)
\ee
be the first-order differential equation to be integrated. Several methods have been developed to do this numerically \cite{numrec}. The simplest one is Euler's method: Imagine that one has estimated the slope at some point $(x_i, y_i)$ by computing $\gamma=f(x_i,y_i)$ from the virial potential-energy fluctuations by means of \eq{gamma}. The point $(x_{i+1}, y_{i+1})$ is then calculated from 

\begin{eqnarray}
x_{i+1}&=&x_i+h \nonumber\\
y_{i+1}&=&y_i + h f(x_i, y_i) + O(h^{2})\,.
\end{eqnarray}
Here, $h$ is the size of the numerical integration step along $x$. The truncation error on the estimated $y_{i+1}$ scales as $h^2$. The statistical error on the numerical calculation of the slope $f$ scales as $1/\sqrt{\tau}$ where $\tau$ is the simulation time. Thus, the statistical error on $y_{i+1}$ scales as $h/\sqrt{\tau}$ (rounding errors from the finite machine precision are not relevant for the $h$'s investigated here). The scaling of the total error is thus proportional to $h^2 + c h/\sqrt{\tau}$ in which $c$ is a constant. We are interested, however, in the ``global'' truncation error, i.e., the accumulated error for some integration length $\Delta x$. Let $N = \Delta x/h$ be the number of steps needed to complete the integration. The total simulation time is $t=N(\tau+\tau_{eq})$ where $\tau_{eq}$ is the time it takes for the system to come into equilibrium when temperature and density are changed. Thus $\tau = t/N - \tau_{eq}$, and with $h=\Delta x/N$ the statistical error on $y$ is $ch/\sqrt{\tau}=c \Delta x/\sqrt{Nt-N^2\tau_{eq}}$. The global error from truncation scales as $N$ since it is systematic, while the statistical error scales as $\sqrt{N}$ due to its randomness. Thus, the total global error is proportional to $(\Delta x)^2/N + c \Delta x/\sqrt{t-N\tau_{eq}}$. The first term is lowered by making $N$ large, while the second term favors small $N$'s and diverges as $N\rightarrow t/\tau_{eq}$. Thus, since $c$ is in general unknown, the optimal choice of $N$ for a given $t$ and $\Delta x$ is not straightforward to determine. We give below a recipe for the optimal parameter choice. First, however, we show how to reduce the truncation error significantly by adopting a higher-order integration method, using the often favored fourth-order Runge-Kutta (RK4) method: For a given point $(x_i,y_i)$, if one defines

\begin{eqnarray}
k_1&=&h f(x_i, y_i)\nonumber\\
k_2&=& h f(x_i+h/2, y_i+k_1/2)\nonumber\\
k_3&=& h f(x_i+h/2, y_i+k_2/2)\\
k_4&=& h f(x_i+h,y_i+k_3)\nonumber\,,
\end{eqnarray}
the next point $(x_{i+1}, y_{i+1})$ is computed as

\begin{eqnarray}
x_{i+1}&=&x_i+h\nonumber\\
y_{i+1}&=& y_i+k_1/6+k_2/3+k_3/3+k_4/6 + O(h^{5})\,.
\end{eqnarray}
While the simple Euler method has a truncation error scaling as $O(h^2)$, the truncation error of RK4 scales as $O(h^5)$. This allows for significantly larger steps along $x$ and thus smaller $N$. From the same type of arguments as given above for the Euler method, the global error of the RK4 method scales approximately as $(\Delta x)^5/N^4 + c\Delta x/\sqrt{t-N\tau_{eq}}$ where $c$ is a (new) unknown constant. 

\begin{figure}[h]\,.
	
	\includegraphics[width=6cm]{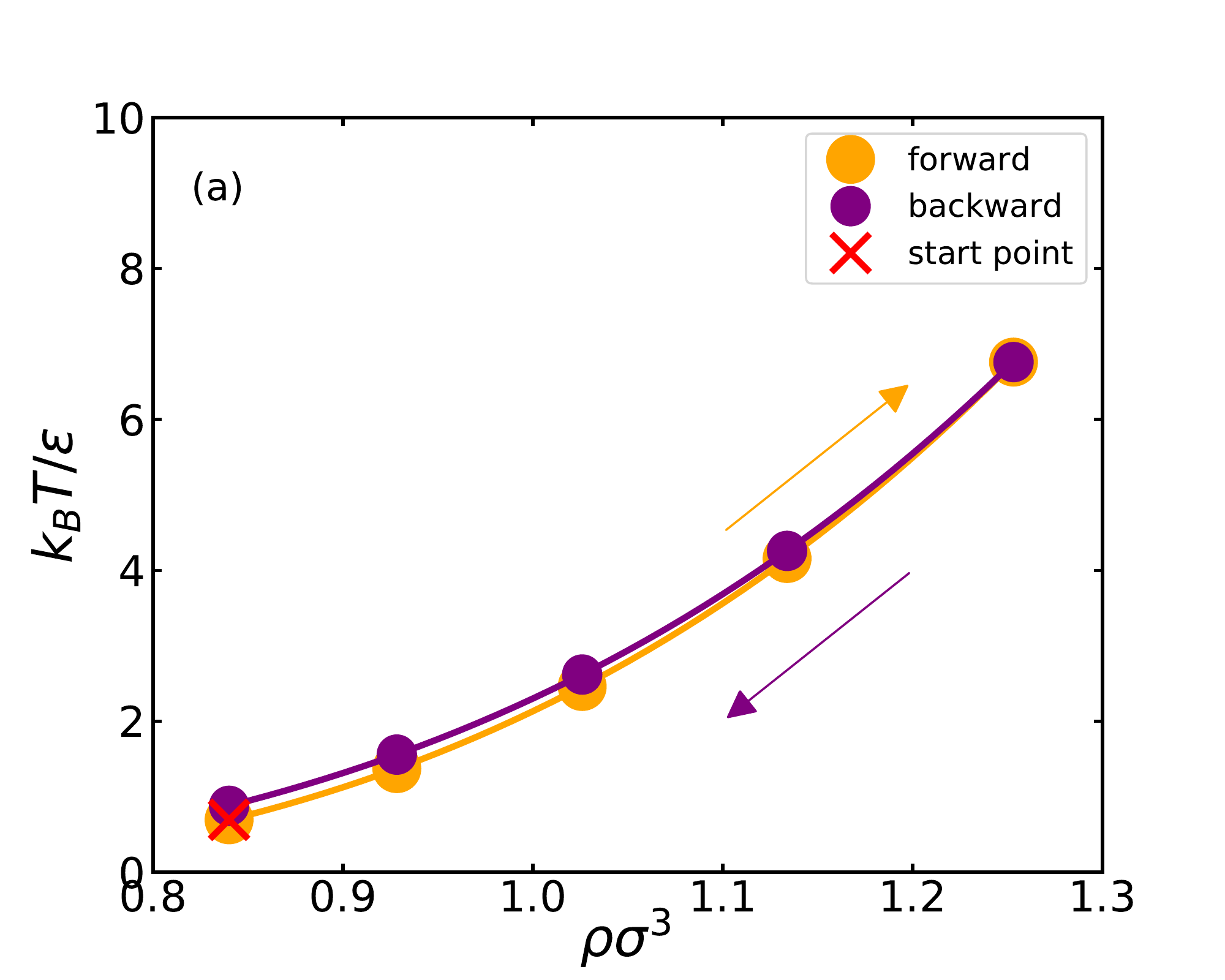} 
	\includegraphics[width=6cm]{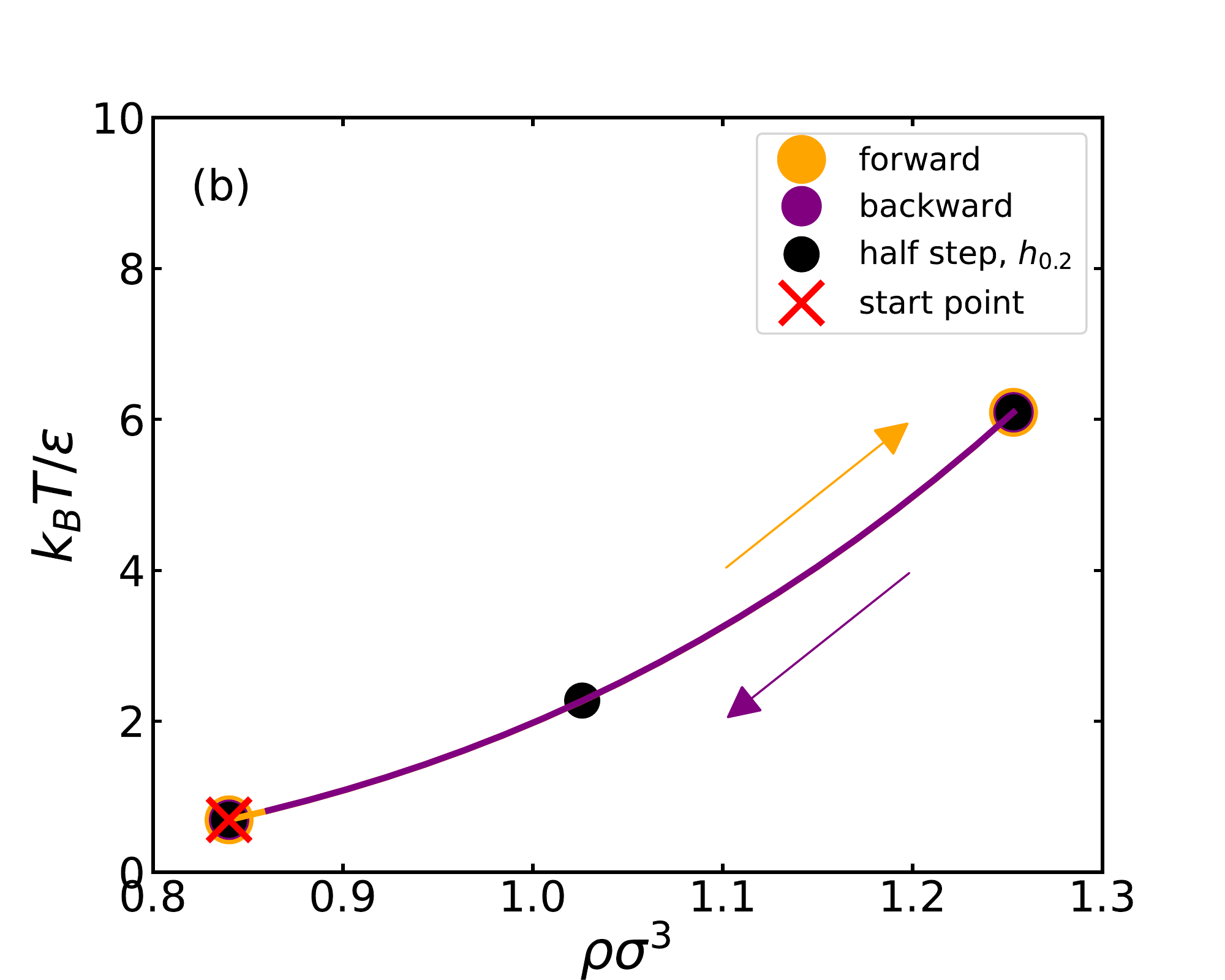}
	\caption{\label{Fig:forward_backward}Configurational adiabat of the WCA system traced out in the thermodynamic phase diagram. (a) The Euler method; (b) the RK4 method. The Euler integration uses a log-density step of size $h=0.1$ (steps in density of $e^{0.1}-1\simeq10$\%), while the RK4 uses $h=0.4$, corresponding to density variation of $e^{0.4}-1\simeq50$\%. The temperature difference of the here presented combined forward-backwards integration $\Delta T$ provides a convenient measure of the maximum error of the predicted temperature. We find $\Delta T\cong 0.186$ for the Euler algorithm and $\Delta T\cong 0.002$ for the RK4 algorithm. The solid lines are interpolations using a cubic Hermite spline.}
\end{figure}

\begin{figure}[h]\,.
	\includegraphics[width=6cm]{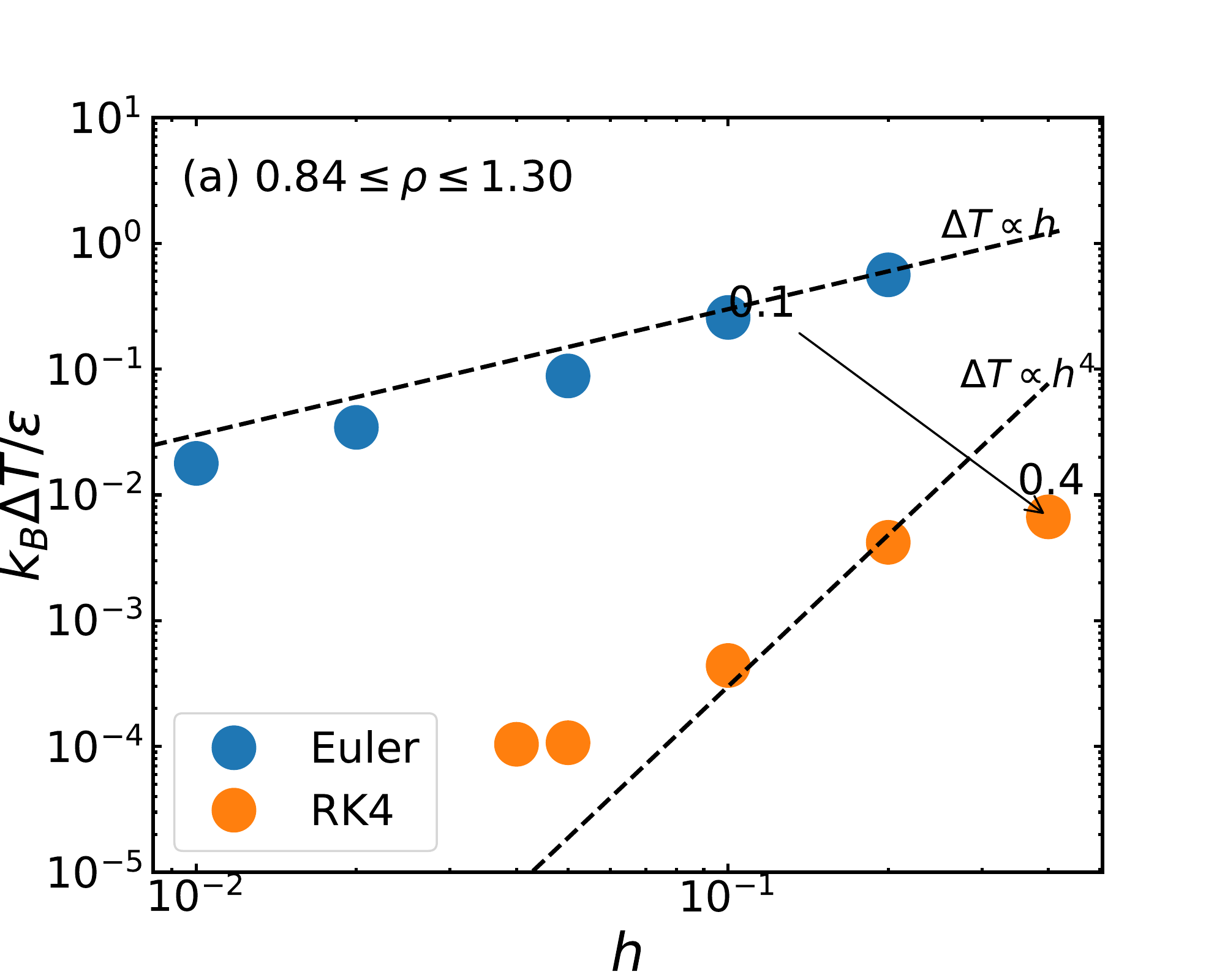}
	\includegraphics[width=6cm]{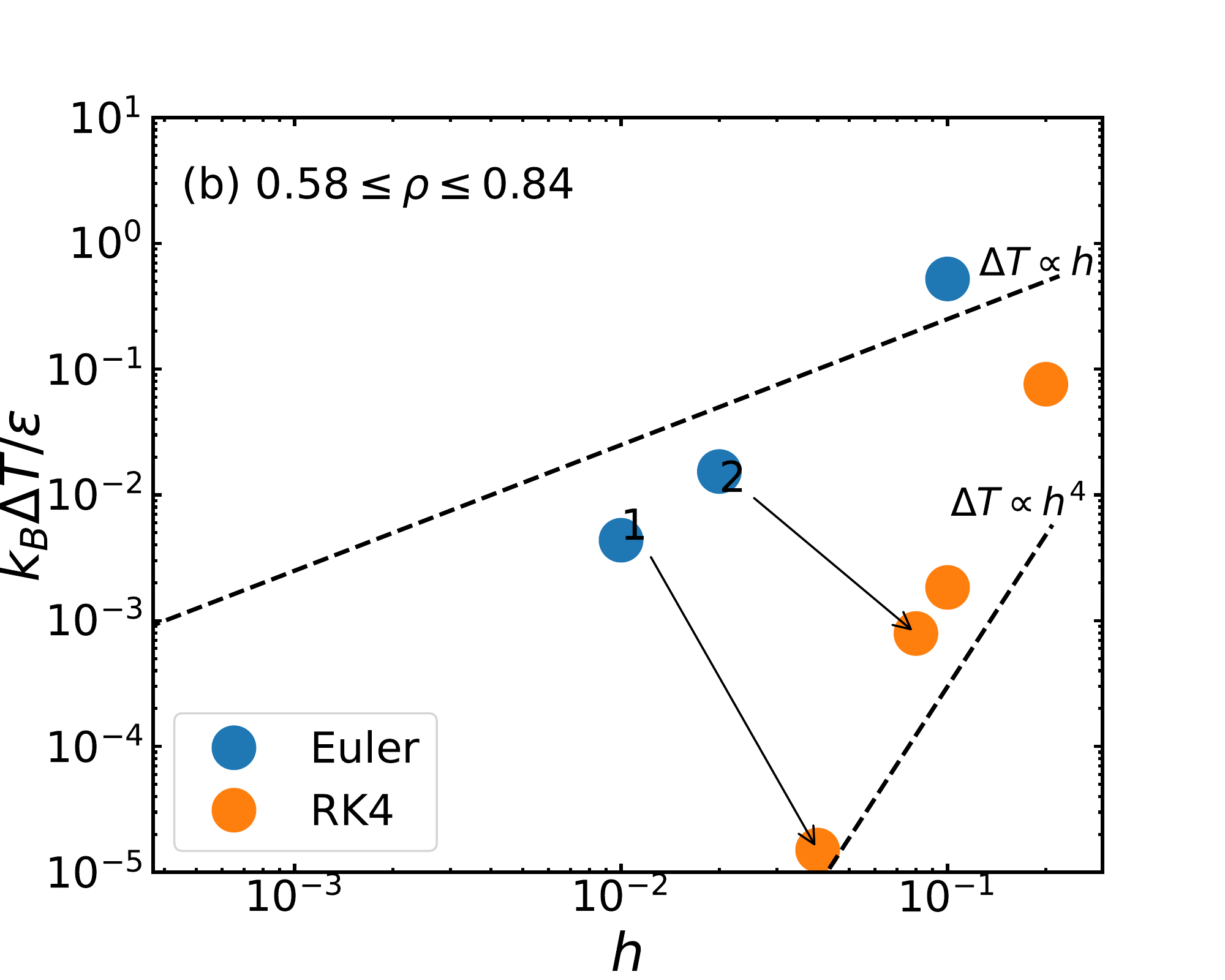} 
	\caption{\label{Fig:step_size}(a) The temperature difference $\Delta T$ of the forward-backward integration in Fig.\ \ref{Fig:forward_backward}, for different steps sizes $h$. The blue dots show results for Euler integration and the orange dots show results for RK4 integration. The temperature difference measures the maximum error in the integration interval $0.84\leq\rho\leq1.30$. The RK4 is significantly more accurate than the Euler algorithm, which allows for larger $h$ steps. The dashed lines indicate the expected scaling of the global error from truncation -- deviations stem from statistical errors on the estimated slopes (slopes are evaluated using simulations lengths of $\tau=655$). The arrow connects Euler and RK4 calculations with approximately the same computational cost (see \fig{Fig:forward_backward}). (b) Same analysis for the integration interval $0.58\leq\rho\leq0.84$.}
\end{figure}

To compare the Euler and RK4 methods, we use each of them for integrating from the initial state point $(\rho, T)=(0.84,0.694)$ to density 1.25 and back again to the initial density of 0.84, see Fig.\ \ref{Fig:forward_backward}. This involves a $\gamma$ variation from 6.825 at the intial density to 4.539 at $\rho=1.25$. The difference between the final temperature of the down integration and the initial temperature, denoted by $\Delta T$, provides a measure of the maximum temperature error. Ideally $\Delta T=0$. Since the RK4 involves four simulations per step, we compare its accuracy where $h$ is four times larger than for the Euler method, which corresponds to approximately the same wall-clock time for the computation. With this constraint, the RK4 is still about two orders of magnitude more accurate: we find $\Delta T=0.186$ for the Euler algorithm and $\Delta T\cong 0.002$ for RK4. Figure \ref{Fig:step_size} shows estimates of the maximum error $\Delta T$ for several values of $h$. To focus on the truncation error, we performed long-time simulations with $\tau\cong 650$. Nonetheless, this analysis demonstrates that a significantly smaller $N$ (larger $h$) is allowed for with the RK4.

Since the RK4 algorithm allows for large $h$, it can be necessary to interpolate in order to identify additional state-points on the isomorph. The solid lines in Fig.\ \ref{Fig:forward_backward} show such interpolations using a cubic Hermite spline: Define $x_{\phi}$ as a point between the two adjacent points $x_{i}$ and $x_{i+1}$, i.e. let $x_{i} \leq x_{\phi} < x_{i+1}$ where $x_\phi=x_i + \phi [x_{i+1}-x_{i}]$ and $ 0 \leq \phi \leq 1$. The interpolated $y_\phi$ is given by the third-degree polynomial: $y_\phi= Ax_{\phi}^{3}+Bx_{\phi}^{2}+Cx_{\phi}+D$ where $y_{\phi} = y_i + [y_{i+1} - y_{i}] [a{\phi}^{3} + b{\phi}^{2}+ c{\phi}]$. For simplicity we introduce the notation $y_{\phi}'=[y_{\phi} - y_{i}]/[y_{i+1} - y_{i}]$ and write the polynomial as $y_{\phi}' = a{\phi}^{3} + b{\phi}^{2}+ c{\phi}$. The coefficients yielding smooth first derivative are $a =  f_{i}' + f_{i+1}'-2$, $b= 3 - 2f_{i}' - f_{i+1}'$, and $c=f_{i}'$ in which $f_{i}' = f_{i} ({x_{i+1}- x_{i}})/({y_{i+1}- y_{i}})$ and $f_{i+1}' = f_{i+1} ({x_{i+1}- x_{i}})/({y_{i+1}- y_{i}})$ are ``reduced'' slopes at the start- and end point, respectively. The $f'$ slopes are given by known $\gamma$'s along the configurational adiabat; thus no extra simulations are needed to evaluate the interpolation.

\begin{figure}
	\includegraphics[width=6cm]{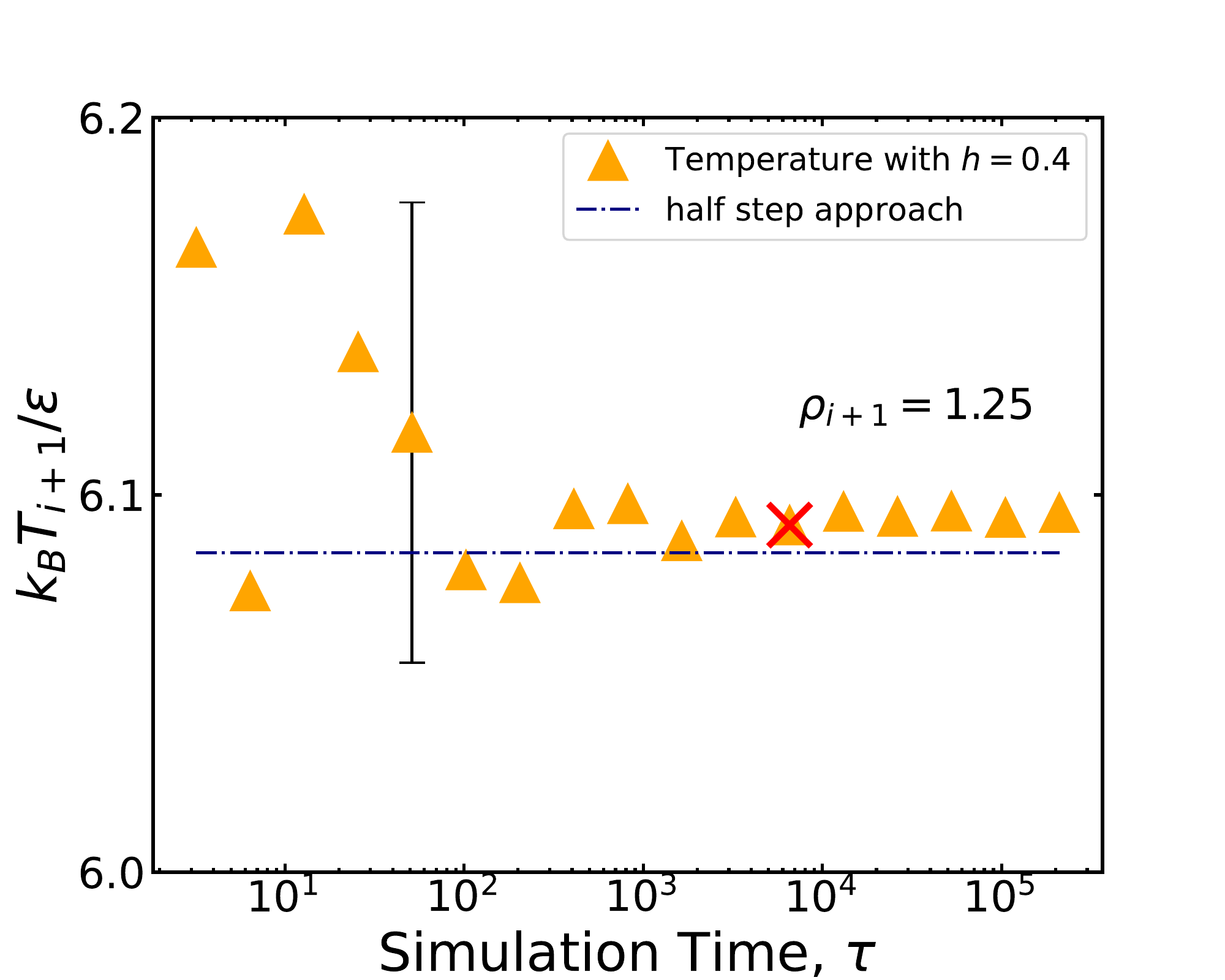}
	\caption{\label{Fig:error}The difference in temperature between using a full step of $h=0.4$ and two half steps of $h=0.2$ when integrating from $\rho=0.84$ up to $\rho=1.25$, plotted against the simulation time per slope evaluation. The desired $h$ can change and the simulation time changes accordingly. The error bar indicates the ``bad statistics with few blocs'' mentioned in the text, computed from Eq. (28) in Ref. \onlinecite{fly89}. The red $\times$ marks the simulation time used in the paper.}
\end{figure}

We investigated the local error by comparing a full $h$ step to two half steps of size $h/2$. The small black dot in the middle of \fig{Fig:forward_backward}(b) shows the results of such two half-steps. The truncation error for the half-step approach is then raised to the sixth order \cite{numrec}, one order higher than RK4 (the price is that one must perform twice as many simulations for each integration step). The triangles on Fig.\ \ref{Fig:error} show the resulting $T_{i+1}$ starting from the reference state-point $(\rho,T)=(0.84,0.694)$, using a full step with $h=0.4$ and varying $\tau$'s. For comparison, the dashed line results from long-time simulations using the half-step algorithm. The distance from triangles to the dashed line provides an estimate of the total error. For short simulation times (small $\tau$'s) the statistical error dominates as seen by the scatter. The truncation error dominates at long simulation times, as seen by the triangles' systematic deviation from the dashed line. For efficient calculation we suggest choosing $h$ and $\tau$ such that the statistical and truncation errors are of the same order of magnitude. The red $\times$ on Fig. \ref{Fig:error} indicates the simulation time $\tau$ used for the figures in the paper.

\begin{figure}
	\includegraphics[width=6cm]{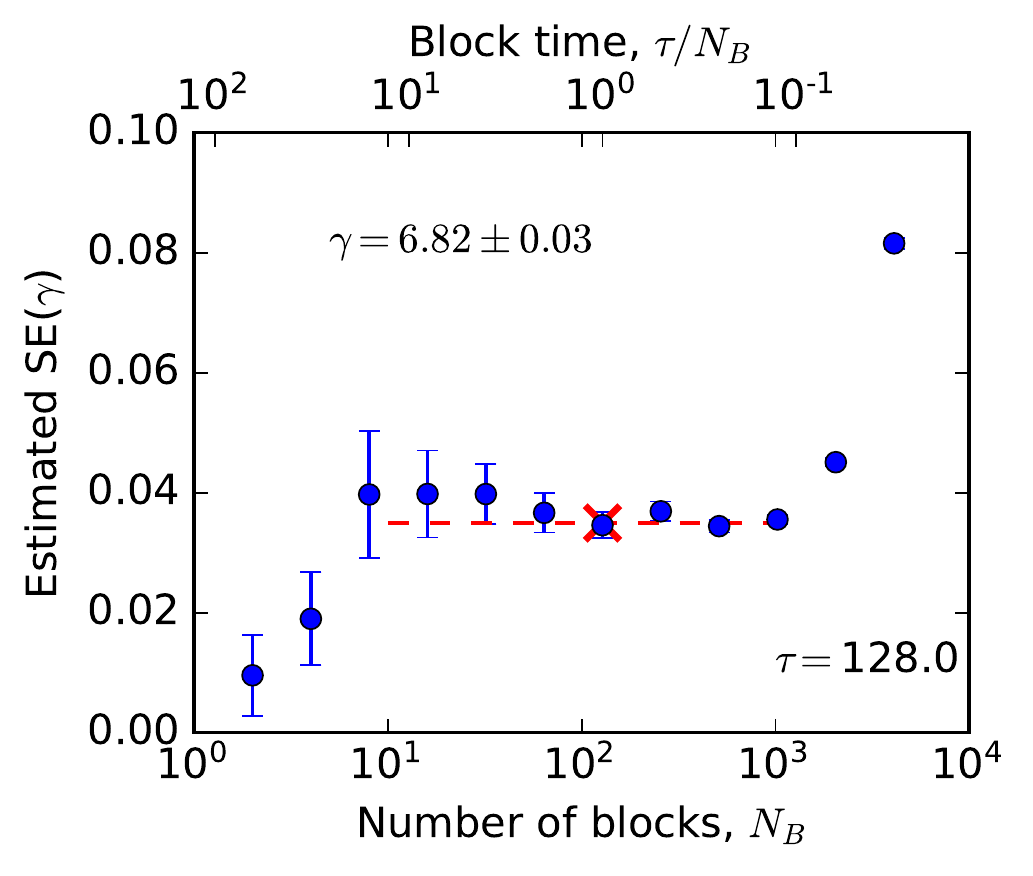}
	\caption{\label{Fig:blocking} Estimate of the statistical error on $\gamma$ from the blocking method. The analysis indicates that $N_B=128$ is an good choice for the number of blocks. This gives $SE(\gamma)=0.03$ on the estimated $\gamma=6.82$.} 
\end{figure}

The above analysis to arrive at the optimal computation time $\tau$ is tedious and involves computationally expensive simulations. We proceed to suggest an efficient optimization recipe that utilizes the fact that the local statistical error of the slopes can be estimated by dividing a given simulation into blocks. If the simulation time for each block is sufficiently long, the blocks are statistically independent. The 67\% confidence standard error is then given by $\text{SE}(\gamma)=\sqrt{\text{VAR}(\gamma)/(N_B-1)}$ where $\text{VAR}(\gamma)$ is the variance of the $\gamma$'s using $N_B$ blocks \cite{fly89}. If the blocks are independent, $\text{VAR}(\gamma)$ scales as $N_B$ and $\text{SE}(\gamma)$ will be independent of the number of blocks. If we divide the simulation into few blocks, $\text{VAR}(\gamma)$ may give a bad estimate of the underlying distribution's theoretical variance. On the other hand, if one divides the simulation into many blocks, the simulation time for each block ($\tau/N_B$) may be brief and the blocks are not independent. In effect, the above formula for $\text{SE}(\gamma)$ gives an overestimate. The optimal $N_B$ is determined by tests of several different $N_B$, as shown on Fig.\ \ref{Fig:blocking} (the red $\times$ corresponds to a good choice of $N_B=128$). The statistical error on $y_{i+1}$ can now be estimated as $\text{SE}(y_{i+1})=\text{SE}(\gamma)h/2$. Here, $2=\sqrt{4}$ enters since the RK4 algorithm includes four independent estimates of slopes (the factor is unity for the Euler algorithm, and $\sqrt{8}$ for the double-step RK4).

Based on the above analysis, we propose the following recipe for efficient and accurate computation of a configurational adiabat:

\begin{enumerate}
	\item Make an $NVT$ simulation at a reference state point of temperature $T_0$ and density $\rho_0$. The simulation time $\tau$ should be sufficiently long that the equilibration time $\tau_{eq}$ can be determined using any standard method (e.g., as the time when the mean-squared displacement has reached the diffusive limit). Use the block method to determine $\text{SE}(\gamma)$, using only the equilibrated part of the trajectory.
	\item Choose $h$. Make a full RK4 step and  estimate the local statistical error using $\text{SE}(y_{i+1})=h\text{SE}(\gamma)/\sqrt{4}$. Use the RK4 two half-step approach to estimate the total local error. If the total local error is unacceptably large, then either
	\begin{enumerate}
		\item increase $\tau$ if the statistical error is of the same magnitude as the total error, or
		\item decrease $h$ if the total error is larger than the statistical error.
	\end{enumerate}
	Small errors suggest that the simulation time, $\tau$, could be decreased or that $h$ can be increased to make the calculation more efficient. $h$ may safely be increased or $\tau$ decreased if the statistical and total errors are of similar magnitude.
	\item Compute adiabatic state-points using the RK4 algorithm with the parameters determined in the above steps. Based on these, a continuous curve can be produced by interpolation using a cubic spline.
	\item Estimate the maximum error by integrating backwards. This error estimate quantifies the accuracy of the computed adiabat.
\end{enumerate}

\begin{figure}
	\includegraphics[width=6cm]{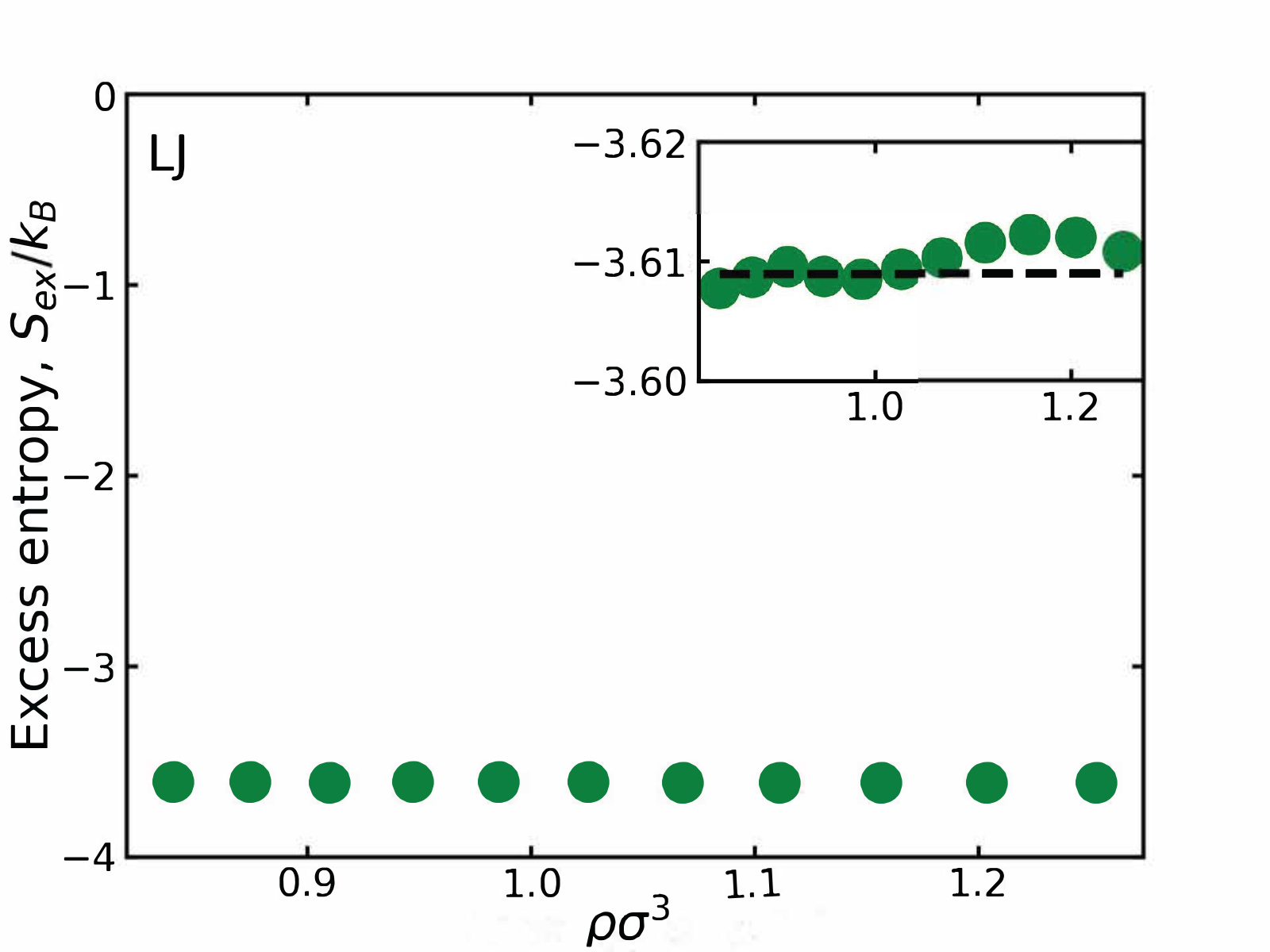} 
	\caption{\label{Fig:sex} The excess entropy values plotted against the densities of the state points on the configurational adiabat traced out for the single-component LJ system starting from the triple point ($\rho=0.84$, $T=0.694$) using RK4 with $h=0.04$. The values are zoomed in to see the deviation from the average value, the black dotted line.}
\end{figure}

As a consistency check of this recipe, \fig{Fig:sex} shows the excess entropy from the equation of state (EOS) of the single-component LJ system by in Ref. \onlinecite{kol94}. The agreement with the configurational adiabat of this EOS is excellent.

\section*{Appendix II: State point data for isomorphs 0-3}

\begin{table}[ht]
	\centering
	\begin{tabular}{rccccccl}
		Selected state points of isomorph 0 (Fig. 1).
	\end{tabular}
	\begin{tabular}{c|c|c|c|c|c|c}
		
		\hline
		$\rho\sigma^{3}$ &$k_BT/\varepsilon$  & $P\sigma^{3}/\varepsilon$ &  $\gamma$ &$R$ & $U/(N\varepsilon)$& $W/(N\varepsilon)$ \\
		\hline
		1.714 & 13.41 & 464.2   & 4.288  & 0.9995  & 50.67&  257.5 \\
		1.636 & 10.98 & 357.7  &  4.329  & 0.9993  &39.92& 207.7 	\\
		1.493 & 7.360  & 211.9  &  4.435  & 0.9988 &24.52 & 134.6 	\\
		1.366 & 4.933  & 125.2 &  4.582 & 0.9978 & 14.84& 86.70  \\
		1.254 & 3.307  & 73.86 &  4.787  & 0.9961  & 8.841& 55.59\\
		1.156 & 2.217  &43.60  &5.068 & 0.9936 & 5.190& 35.49 \\  
		1.071 & 1.486 & 25.81 & 5.445 & 0.9902 & 3.007 & 22.60 	\\		
		0.9985 & 0.9960 & 15.35 &  5.939 & 0.9860 & 1.721 & 14.38\\			
		0.9364 & 0.6677 & 9.200 &  6.571 & 0.9808 & 1.298& 9.157\\
		0.9091 & 0.5466 & 7.145 &  6.945 & 0.9782 & 0.7332& 7.313 	\\
		0.8610 & 0.3664 & 4.341 & 7.835 & 0.9726 &0.4117 &4.675  \\			
		0.8400 & 0.3000 & 3.396 &  8.353 & 0.9698 &0.3079& 3.743 	\\
		0.8207 & 0.2456 & 2.664 & 8.932 & 0.9671& 0.2300 &2.100 \\
		0.7592 & 0.1104 & 1.034 & 11.94 & 0.9566& 0.0711& 1.251 \\
		0.7168 & 0.04960 & 0.4159 &  16.42 & 0.9475&0.0218 &0.5306 \\
		0.6877 & 0.02230  &0.1725  & 23.09 & 0.9402&0.006653& 0.2285 \\
		0.6680 & 0.009059 & 0.06587 & 33.06 & 0.9349 &0.001742 &0.08984 \\
		0.6546 & 0.004972 & 0.03509 &47.83 & 0.9304& 0.0007115 & 0.04379 \\
		0.6456 & 0.002021 & 0.01382 &  69.90 & 0.9277& 0.0001853 & 0.01938\\
		0.6353 &0.0004081 & 0.002703 &152.0 & 0.9243 &0.00001690 & 0.003846  \\
		\hline
	\end{tabular}		
\end{table}	

\begin{table}[ht]
	\centering
	\begin{tabular}{rccccccl}
		Selected state points of isomorph 1 (Fig. 1).
	\end{tabular}
	\centering
	\begin{tabular}{c|c|c|c|c|c|c}
		\hline
	$\rho\sigma^{3}$ &$k_BT/\varepsilon$  & $P\sigma^{3}/\varepsilon$ &  $\gamma$ &$R$ & $U/(N\varepsilon)$& $W/(N\varepsilon)$ \\
	\hline	
		
		1.565 & 14.72 & 340.4 & 4.337 & 0.9993&  39.48&202.9\\
		1.495 & 12.05 & 262.7  &4.385  &0.9991 & 31.11& 163.7\\
		1.366 & 8.078 & 156.1 &  4.506 & 0.9983 & 19.14& 106.2\\
		1.252 & 5.415 & 92.69 &  4.671  &0.9971 & 11.61& 68.63\\
		1.151 & 3.630 & 55.03 &  4.893  &0.9954 & 6.956&44.18\\
		1.024 & 1.992 & 25.27 &  5.366 & 0.9913 & 3.149 &22.70\\
		0.9527 & 1.335  &15.12 &  5.799 & 0.9875 & 1.830 &14.53\\
		0.8916 & 0.8951 & 9.101 & 6.351 & 0.9831 & 1.053 & 9.310\\
		0.8400& 0.6000 & 5.520 & 7.041 &  0.9782 & 0.6015 &5.972 	\\
		0.7961 & 0.4022  &3.377 &  7.899 & 0.9730& 0.3412&3.840\\
		0.7590 & 0.2696 & 2.085 &  8.955 & 0.9677& 0.1925&2.477\\
		0.7280 & 0.1807 & 1.299 & 10.25& 0.9624& 0.1081 & 1.603 \\
		0.7019 & 0.1211 & 0.8161 & 11.84 & 0.9574 &0.06048 & 1.041\\
		0.6708 & 0.06648 & 0.4129 & 14.88 & 0.9504 & 0.02517&0.5491\\
		0.6543 & 0.04456 & 0.2646 & 17.48 & 0.9465 & 0.01399&0.3598 \\
		0.6245 & 0.01639&  0.08940 &  26.75 & 0.9380 &  0.003198 &0.1268\\
		0.6060 & 0.006031 & 0.03111 & 42.01 & 0.9319 &0.0007245&0.04532\\
		0.5945 & 0.002219 & 0.01105 & 67.19 & 0.9282& 0.0001632& 0.01637\\
		0.5787 & 0.00002724 & 0.0001290 &  579.6 & 0.9222 &0.0000002251&0.0001957 \\
			\hline
	\end{tabular}
\end{table}

\begin{table}[ht]
	\centering
	\begin{tabular}{rccccccl}
		Selected state points of isomorph 2 (Fig. 1).		
	\end{tabular}
	\centering
	\begin{tabular}{c|c|c|c|c|c|c}
	\hline
	$\rho\sigma^{3}$ &$k_BT/\varepsilon$  & $P\sigma^{3}/\varepsilon$ &  $\gamma$ &$R$ & $U/(N\varepsilon)$& $W/(N\varepsilon)$ \\
	\hline	
	1.403 & 13.46  &219.9 &  4.415 & 0.9989& 27.20&143.3\\
	1.341 & 11.02 & 169.8 &  4.474 & 0.9985&  21.39 &115.6\\
	1.228 & 7.389 & 101.2 & 4.620 & 0.9976& 13.12 &75.04 \\
	1.128 & 4.953 & 60.34&   4.814 & 0.9961& 7.946&48.53\\
	1.040 & 3.320 & 36.01 & 5.070& 0.9940& 4.756 &31.30 \\
	1.001 & 2.718 & 27.85 & 5.225 & 0.9926& 3.664 &25.11 \\
	0.9637 & 2.226 & 21.56 &  5.399 & 0.9919& 2.814&20.14\\
	0.8972 & 1.492  &12.96 & 5.820 & 0.9876& 1.648&12.95 \\
	0.8675& 1.221 & 10.07 &  6.071 &  0.9856&1.256 &10.39 \\
	0.8400 & 1.000 & 7.837 &  6.350 & 0.9834& 0.9557&8.330\\
	0.8146& 0.8187 & 6.110 & 6.663 & 0.9811 &  0.7256& 6.683 \\
	0.7494& 0.4493& 2.930 & 7.828 & 0.9737 &0.3141&3.461\\
	0.6987 & 0.2466& 1.432 & 9.411 & 0.9659 & 0.1342&1.803 \\
	0.6295 & 0.07427 & 0.3613 & 14.45 & 0.9515&0.02378 &0.4998 \\
	0.5945 & 0.02734 & 0.1204 & 21.64 & 0.9420& 0.005505 &0.1752\\
	0.5694 & 0.008230 & 0.03360 &  36.68 & 0.9337 & 0.0009355	&0.05084 \\
	0.5591 & 0.003698 & 0.01464& 53.01 & 0.9300& 0.0002851 &0.02248\\
	0.5507 & 0.001360& 0.005242 & 85.02 & 0.9268& 0.00006423 &0.008158 \\
	0.5436 & 0.0002747 & 0.001034&  185.1 &0.9238& 0.000005878 & 0.001628 \\
	\hline
	\end{tabular}
\end{table}

\begin{table}[ht]
	\centering
	\begin{tabular}{rccccccl}
		Selected state points of isomorph 3 (Fig. 1).		
	\end{tabular}
	\centering
	\begin{tabular}{c|c|c|c|c|c|c}
	\hline
$\rho\sigma^{3}$ &$k_BT/\varepsilon$  & $P\sigma^{3}/\varepsilon$ &  $\gamma$ &$R$ & $U/(N\varepsilon)$& $W/(N\varepsilon)$ \\
\hline	
		1.261 & 14.79& 160.7 & 4.468 & 0.9986&21.25& 112.7\\
		1.206 & 12.10 & 124.4 & 4.531 & 0.9982&91.04 & 16.73\\
		1.106 & 8.110 & 74.47 &  4.687 & 0.9971& 10.30&59.25\\
		1.060 & 6.640 & 57.64 & 4.781  &0.996& 8.044 &47.75 \\
		0.9766 & 4.451 & 34.57 & 5.011 & 0.9946 & 4.870&30.95 \\
		0.9389&  3.644 & 26.80 & 5.148 &  0.9934& 3.774&24.90 \\
		0.9036 & 2.984 & 20.79 & 5.304 & 0.9922 & 2.917&20.03 \\
		0.8400 & 2.000 & 12.55 & 5.675 & 0.9890& 1.730& 12.94\\
		0.8114 & 1.638 & 9.771 &  5.892 & 0.9873 & 1.327& 10.41 \\
		0.7603 & 1.098 & 5.947 &  6.410 & 0.9833& 0.7760	&6.725 \\
		0.6787 & 0.4932  & 2.248 &  7.836 & 0.9741& 0.2592 &2.820 \\
		0.6196& 0.2216 & 0.8762 &  9.977 & 0.9641 & 0.08440 &  1.192 \\
		0.5776 & 0.0996 &  0.3520 &  13.15  & 0.9548 & 0.02690&0.5098\\
		0.5481 &0.04474 & 0.1453 &  17.89  & 0.9466 & 0.008466 &0.2204\\
		0.5276 & 0.02010 & 0.06135 &  24.92 & 0.9395& 0.002630 &0.09617 \\
		0.5135 & 0.009033 & 0.02636 &  35.42 & 0.9347&  0.0008093& 0.04231 \\
		0.5038 & 0.004059 & 0.01148 &  50.9571 & 0.9305& 0.0002475 &0.01873 \\
		0.4973 & 0.001824 & 0.005048 &  74.1566 & 0.9278  & 0.00007535 & 0.008328\\
		0.4920 & 0.0006709 & 0.001824 &  119.9858 & 0.9254 & 0.00001656 &0.003037 \\
		\hline
	\end{tabular}
\end{table}

\end{document}